# Silicon-plasmonic integrated circuits for terahertz signal generation and coherent detection

T. Harter[1,2*], S. Muehlbrandt[1,2], S. Ummethala[1,2], A. Schmid[1], L. Hahn[2], W. Freude[1], C. Koos[1,2**]

[1]Institute of Photonics and Quantum Electronics (IPQ), Karlsruhe Institute of Technology (KIT), 76131 Karlsruhe, Germany
[2]Institute of Microstructure Technology (IMT), Karlsruhe Institute of Technology (KIT), 76131 Karlsruhe, Germany
*tobias.harter@kit.edu, **christian.koos@kit.edu

**Optoelectronic signal processing offers great potential for generation and detection of ultra-broadband waveforms in the THz range, so-called T-waves. However, fabrication of the underlying high-speed photodiodes and photoconductors still relies on complex processes using dedicated III-V semiconductor substrates. This severely limits the application potential of current T-wave transmitters and receivers, in particular when it comes to highly integrated systems that combine photonic signal processing with optoelectronic conversion to THz frequencies. In this paper, we demonstrate that these limitations can be overcome by plasmonic internal photoemission detectors (PIPED). PIPED can be realized on the silicon photonic platform and hence allow to leverage the enormous opportunities of the associated device portfolio. In our experiments, we demonstrate both T-wave signal generation and coherent detection at frequencies of up to 1 THz. To proof the viability of our concept, we monolithically integrate a PIPED transmitter and a PIPED receiver on a common silicon photonic chip and use them for measuring the complex transfer impedance of an integrated T-wave device.**

Terahertz signals (T-waves) offer promising perspectives for a wide variety of applications, comprising high-speed communications[1–3], microwave photonics[4], spectroscopy[5,6], life sciences[7,8], as well as industrial metrology[9,10]. Optoelectronic signal processing techniques are particularly attractive both for T-wave generation[1,11,12] and detection[13–15], especially when broadband operation is required. On a conceptual level, optoelectronic T-wave generation relies on mixing of two optical signals oscillating at frequencies $f_a$ and $f_b$ in a high-speed photodetector, for which the photocurrent depends on the incident optical power[11]. The photocurrent oscillates with a difference frequency $f_{THz} = |f_a - f_b|$ in the terahertz region, which can be relatively easily adjusted over the full bandwidth of the photodetector by frequency-tuning of one of the two lasers. In many practical applications, the optical signal oscillating at $f_a$ carries an amplitude or phase modulation, whereas the optical signal at $f_b$ is simply a continuous-wave (CW) carrier. In this case, the phase and amplitude modulation of the optical carrier is directly transferred to the T-wave carrier. This concept shows great potential for high-speed wireless communications at THz carrier frequencies and has been at the heart of a series of transmission experiments, in which record-high data rates of 100 Gbit/s and above have been reached[2,16–18]. Similarly, optoelectronic techniques can be used for detection of T-wave signals. In this case, the T-wave signal is applied to a high-speed photoconductor and the optical power oscillation at the difference frequency $f_{THz} = |f_a - f_b|$ is used as a local oscillator for coherent down-conversion to the baseband[13–15]. This technique is mainly used in frequency-domain THz spectroscopy systems offering a widely tunable frequency range and a high signal-to-noise ratio[13,19,20].

For exploiting the tremendous application potential of optoelectronic T-wave processing, monolithic co-integration of photonic devices and T-wave transmitters and receivers are of vital importance. From the technology side, however, optoelectronic T-wave transmitters and receivers are still rather complex, relying on high-speed photodiodes[21–23] or photoconductors[13,24] that require dedicated III-V semiconductor substrates that are obtained, e. g., through low-temperature growth of InGaAs/InAlAs multilayer structures[25] and that are not amenable to large-scale photonic integration. This does not only hamper the co-integration of T-wave transmitter and receiver circuitry on a common chip but also hinders the exploitation of highly developed photonic integration platforms for building advanced optoelectronic T-wave systems that combine photonic signal processing with optoelectronic frequency conversion on a common chip.

In this paper, we demonstrate an approach that allows to integrate T-wave transmitters and receivers directly on the silicon photonic platform, thereby exploiting the outstanding technical maturity, scalability and the comprehensive device portfolio[26-28] of this material system. The approach exploits internal photoemission at metal-semiconductor interfaces of plasmonic structures[29,30] that can be directly integrated into widely used silicon-on-insulator (SOI) waveguides. Our experiments show that these plasmonic internal photoemission detectors (PIPED) are not only suited for photomixing in the T-wave transmitter, but also lend themselves to highly sensitive optoelectronic reception. In a proof-of-concept experiment, we monolithically co-integrate a PIPED transmitter and a PIPED receiver on a common silicon photonic chip and use them for measuring the complex transfer function of an integrated T-wave transmission line. In this context, we also develop and experimentally verify a mathematical model of optoelectronic T-wave conversion that allows to quantitatively describe T-wave generation and detection over a wide range of frequencies.

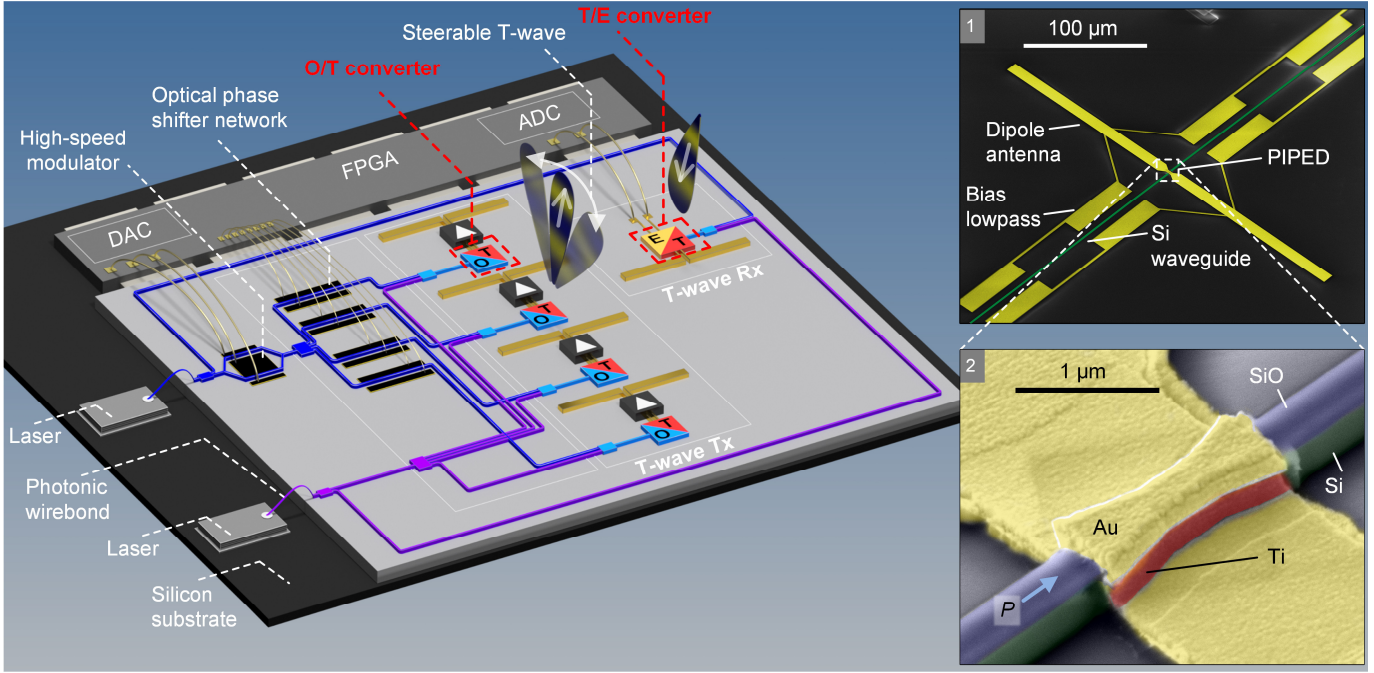

**Figure 1** Vision of an integrated silicon-plasmonic T-wave wireless transceiver that exploits optoelectronic signal processing both at the transmitter and at the receiver. The system combines optical-to-T-wave (O/T) and T-wave-to-electronic (T/E) converters with advanced silicon photonic devices such as phase shifters or high-performance modulators. Continuous-wave (CW) lasers are coupled to the chip using photonic wire bonds[38], and electrical circuits such as field-programmable gate arrays (FPGA), digital-to-analogue and analogue-to-digital converters (DAC and ADC) are used to drive the modulators and to further process the received signals. The transmitter comprises a T-wave antenna array, fed by an array of O/T converters that are driven by a series of optical signals. The phases of the optical and hence the T-wave signals can be precisely defined by an electrically driven optical phase shifter network, thereby enabling broadband beam steering. Optionally, additional T-wave amplifiers could boost the power of the generated signals. Coherent detection of the T-wave signal relies on a pair of optical carriers, the power beat of which serves as a local oscillator for T/E conversion. The monolithic co-integration of optical phase shifter networks, T-wave transmitters and T-wave receivers on a single chip paves the route to highly stable, compact and inexpensive THz systems. **Inset 1**: For O/T and T/E conversion, the concept relies on plasmonic internal-photoemission detectors (PIPED[29]) that are coupled to dipole antennae. The PIPED are fed through silicon photonic waveguides and biased via low-pass structures that are directly connected to the arms of the dipole antennae. **Inset 2**: Detailed view of a fabricated PIPED. The device consists of a narrow silicon nanowire waveguide that is combined with overlays of gold (Au) and titanium (Ti) to form an ultra-small plasmonic structure with two metal-semiconductor interfaces. The optical power $P$ is fed to the PIPED with by a silicon photonic waveguide.

**Silicon-plasmonic T-wave systems**

The vision of an integrated silicon-plasmonic T-wave system is illustrated in Fig. 1 using a wireless high-speed transceiver as an exemplary application case. The system combines T-wave transmitter, T-wave receiver and a variety of other silicon photonic devices[26,27] such as phase shifters[31–33] or high-performance modulators[34–37] on a common substrate. Continuous-wave (CW) lasers are coupled to the chip using photonic wire bonds[38], and electrical circuits such as field-programmable gate arrays (FPGA), digital-to-analogue and analogue-to-digital converters (DAC and ADC) are used to drive the modulators and to further process the received signals. T-wave generation is accomplished by photomixing of modulated optical signals with an optical CW tone in high-speed plasmonic internal-photoemission detectors (PIPED[29]). The antenna-coupled PIPED act as optical-to-T-wave (O/T) converters. Large-scale monolithic integration of advanced silicon photonic devices with O/T converters opens rich opportunities for advanced T-wave signal processing. In the illustration, the transmitter comprises a T-wave antenna array, fed by an array of O/T converters that are driven by a series of optical signals. The phases of the optical signals and hence those of the T-wave signals can be precisely defined by an electrically driven optical phase shifter network[31,32], thereby enabling broadband beam steering and shaping. Optionally, integrated T-wave amplifiers can be used to boost the T-wave signals. At the receiver, optoelectronic down-conversion (T/E-conversion) is used for coherent detection of the T-wave signal, using the power beat of two optical waves as a local oscillator. For O/T and T/E conversion, the concept relies on PIPED that are coupled to dipole antennae, see Inset 1 of Fig. 1. The PIPED are fed through silicon photonic waveguides and biased via dedicated low-pass structures that are directly connected to the arms of the dipole antennae. A more detailed view of a PIPED is shown in Inset 2. The device consists of a narrow silicon nanowire waveguide that is combined with overlays of gold (Au) and titanium (Ti) to form an ultra-small plasmonic structure with two metal-semiconductor interfaces. Note that the PIPED concept does not rely on the use of gold as a plasmonic material – this was chosen only for ease of fabrication in the current experiment. When combined with large-scale silicon photonic circuits, gold-free designs may be used that allow the processing in a state-of-the-art CMOS line[39].

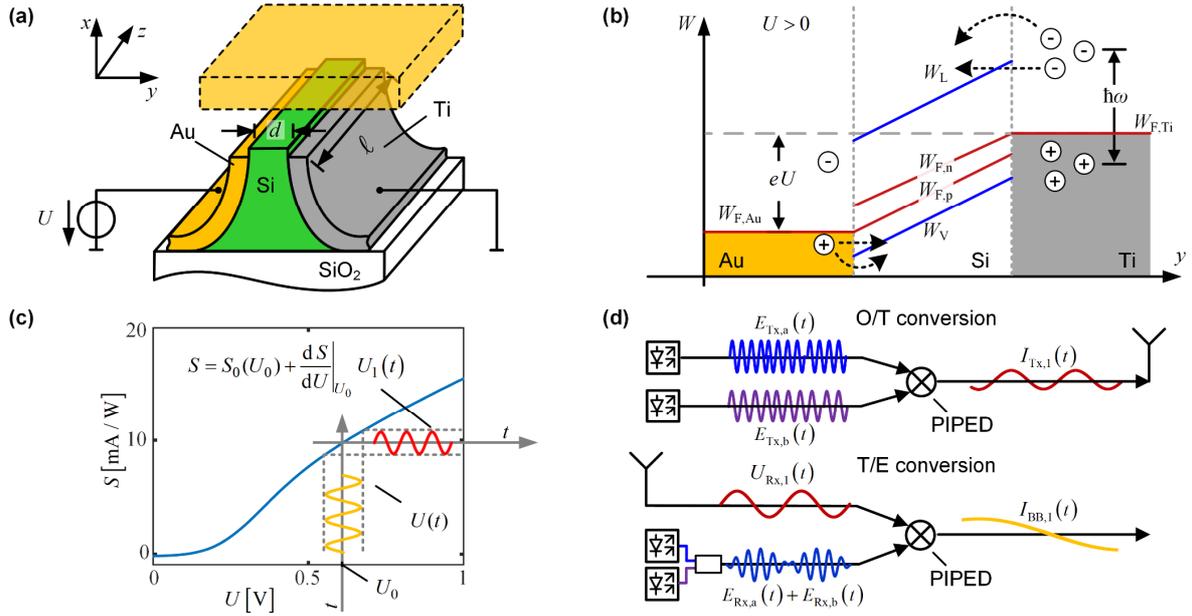

**Figure 2** Operating principle of THz transmitter and coherent receiver using a plasmonic internal-photoemission detector (PIPED) on a silicon chip. **(a)** Schematic of the PIPED, consisting of a silicon core with a gold and a titanium sidewall. Surface plasmon polaritons (SPP) are guided along the z-direction, and are mostly absorbed in the Ti due to the large imaginary part of its complex electric permittivity $\underline{\varepsilon}_r$. A voltage $U$ is applied between the Au and the Ti electrode. **(b)** Band diagram of the Au-Si-Ti junction when a voltage $U > 0$ is applied between the electrodes. Photons absorbed in the Ti layer excite hot electrons, which can overcome or tunnel through the Schottky potential barrier, leading to a photocurrent $I = SP$ in proportion to the number of absorbed photons, i. e., in proportion to the optical power $P$. An equivalent effect occurs for holes at the Au-Si-interface – the relative magnitude of the two contributions is still under investigation. The carrier emission probability into the Si waveguide core and therefore the sensitivity $S(U)$ can be increased ($U > 0$) or decreased ($U < 0$) by varying the bias voltage $U$. **(c)** Measured sensitivity $S(U)$ of the PIPED in dependence of the applied voltage $U$. The sensitivity $S(U)$ can be linearized in the vicinity of a bias voltage $U = U_0$. **(d)** In essence, the PIPED can be used as a mixer that multiplies two signals to generate a waveform at the difference-frequency. **Upper Subfigure:** When used for photomixing (O/T conversion) at the transmitter (Tx), the PIPED acts as a quadratic detector and the output photocurrent $I_{Tx,1}(t)$ corresponds to the product of two time-dependent optical signals $E_{Tx,a}(t)$ and $E_{Tx,b}(t)$. **Lower Subfigure:** When used for optoelectronic downconversion (T/E conversion) at the T-wave receiver (Rx), the PIPED combines two functionalities, namely the generation of a THz local oscillator (LO) from two optical carriers $E_{Rx,a}(t)$ and $E_{Rx,b}(t)$ and the down-conversion of the received THz signal to the baseband. To this end, the PIPED is fed by a superposition of two unmodulated optical carriers, oscillating at frequenceis $f_{Rx,a}$ and $f_{Rx,b}$, while a time-dependent voltage $U_{Rx,1}(t)$ modulates the device sensitivity. The PIPED photocurrent is then given by the product of the time-variant sensitivity with the time-variant optical power $P_{Rx}(t)$.

**Plasmonic internal photoemission detectors (PIPED) for optoelectronic T-wave processing**

The PIPED concept is illustrated and explained in Fig 2. Figure 2(a) shows a schematic cross section of the device. The Si nanowire waveguide core is contacted by an Au layer on the left and by a Ti layer on the right – details of the fabrication can be found in ref. 29. To drive the device, light at infrared telecommunication wavelengths ($\lambda = 1.5\ \mu m$) is coupled to the Si waveguide core, leading to an excitation of surface plasmon polaritons (SPP) both at the Au-Si and at the Si-Ti interfaces – the associated energy levels are sketched in Fig. 2(b) for a forward bias voltage U > 0, which is counted positive from the Au to the Ti electrode. Free-carrier absorption generates hot electrons in the titanium with carrier energies above the Fermi level $W_{F,Ti}$. An equivalent effect occurs for holes at the Au-Si-interface – the relative magnitude of the two contributions is currently under investigation. The hot electrons and holes have an increased probability to cross the $d = 100$ nm wide Si barrier, leading to a photocurrent $I$ from the Au to the Ti side. The photocurrent $I$ depends linearly on the optical power $P$ with a sensitivity (or responsivity) $S = I/P$. Note that, in contrast to conventional photodiodes, the measured sensitivity $S(U)$ of the PIPED depends on the applied voltage $U$ as shown in Fig 2(c). This is a key aspect for efficient optoelectronic T/E reception. For a forward bias $U > 0$, the band edges tilt inside the Si core, Fig 2(b), which reduces the effective width of the potential barrier such that the sensitivity $S$ increases with $U$. For a reverse bias $U < 0$, the carrier emission probability is small, and hence the photocurrent remains small. The strong attenuation of the SPP allows junction lengths $\ell$ of less than $1\ \mu m$ and device capacitances smaller than $C = 1\ fF$ (ref. 29). With a load resistance of $R = 50\ \Omega$, this would lead to an $RC$ limiting frequency of $3\ THz$. In the current device designs, this limitation is not relevant since the speed is limited by the transit times, for which we estimate 1 ps for electrons and 1.5 ps for holes. This estimation is based on a voltage drop of $U = 0.5\,\text{V}$ at the 100 nm wide barrier, leading to drift velocities of $10^7\,\text{cm}\,\text{s}^{-1}$ for electrons and $6.5 \cdot 10^5\,\text{cm}\,\text{s}^{-1}$ for holes[40], close to respective saturation velocity. This limits the bandwidth to approximately 0.44 THz assuming dominating electron transport and to 0.29 THz in case hole transport dominates[41]. The fast device response makes the PIPED an excellent candidate for T-wave generation and reception at frequencies up to 1 THz and above.

For O/T-conversion, the transmitter (Tx) essentially acts as a mixer that multiplies two time-dependent optical signals $E_{\text{Tx,a}}(t)$ and $E_{\text{Tx,b}}(t)$ to produce a photocurrent $I_{\text{Tx}}(t)$ that corresponds to the difference-frequency waveform, see upper part of Fig. 2(d). In the following, we only give a short mathematical description of photomixing and optoelectronic down-conversion in the PIPED. A rigorous analysis can be found in the Supplementary Section 1. We assume that the optical signal $E_{\text{Tx,a}}(t)$ oscillates at angular frequency $\omega_{\text{Tx,a}}$ and carries an amplitude modulation $\hat{E}_{\text{Tx,a}}(t)$ and/or a phase modulation $\varphi_{\text{Tx,a}}(t)$, whereas the optical signal $E_{\text{Tx,b}}(t)$ is simply a CW carrier with constant amplitude $\hat{E}_{\text{Tx,b}}$, frequency $\omega_{\text{Tx,b}}$ and phase $\varphi_{\text{Tx,b}}$,

$$E_{\text{Tx,a}}(t) = \hat{E}_{\text{Tx,a}}(t)\cos(\omega_{\text{Tx,a}}t + \varphi_{\text{Tx,a}}(t)) \quad (1)$$
$$E_{\text{Tx,b}}(t) = \hat{E}_{\text{Tx,b}}\cos(\omega_{\text{Tx,b}}t + \varphi_{\text{Tx,b}}).$$

The optical power $P_{\text{Tx}}(t)$ then oscillates with the difference frequency $\omega_{\text{Tx,THz}} = |\omega_{\text{Tx,a}} - \omega_{\text{Tx,b}}|$,

$$P_{\text{Tx,1}}(t) = \hat{P}_{\text{Tx,1}}(t)\cos(\omega_{\text{Tx,THz}}t + \varphi_{\text{Tx,THz}}(t)) \quad (2)$$

where the amplitude $\hat{P}_{\text{Tx,1}}(t)$ and the phase $\varphi_{\text{Tx,THz}}(t)$ of the oscillation are directly linked to the normalized amplitude and to the phase of the optical wave,

$$\hat{P}_{\text{Tx,1}}(t) = \hat{E}_{\text{Tx,a}}(t)\hat{E}_{\text{Tx,b}}, \quad \varphi_{\text{Tx,THz}}(t) = \varphi_{\text{Tx,a}}(t) - \varphi_{\text{Tx,b}}. \quad (3)$$

When detected by the PIPED (sensitivity $S_{\text{Tx}}$) this leads to an oscillating component in the photocurrent $I_{\text{Tx}}(t)$, featuring the same frequency and the same phase as the optical power oscillation,

$$I_{\text{Tx,1}}(t) = \hat{I}_{\text{Tx,1}}(t)\cos(\omega_{\text{Tx,THz}}t + \varphi_{\text{Tx,THz}}(t)), \quad (4)$$

where

$$\hat{I}_{\text{Tx,1}}(t) = S_{\text{Tx}}\hat{E}_{\text{Tx,a}}(t)\hat{E}_{\text{Tx,b}}. \quad (5)$$

Hence, any modulation of the amplitude $\hat{E}_{\text{Tx,a}}(t)$ or the phase $\varphi_{\text{Tx,a}}(t)$ of the optical signal translates directly into an amplitude and phase modulation of the THz wave. Exploiting this concept, broadband high-quality THz signals can be generated by using widely available optical communication equipment. The THz field is radiated by an antenna or coupled to a transmission line.

Similarly, PIPED can be used for T/E conversion in the T-wave receiver (Rx), see lower part of Fig. 2(d). In this case, the device combines two functionalities, namely the generation of a THz local oscillator (LO) from two optical carriers, and the down-conversion of the received THz signal to the baseband. To this end, the PIPED is fed by a superposition of two un-modulated optical tones, oscillating at frequencies $\omega_{\text{Rx,a}}$ and $\omega_{\text{Rx,b}}$,

$$E_{\text{Rx,a}}(t) = \hat{E}_{\text{Rx,a}}\cos(\omega_{\text{Rx,a}}t + \varphi_{\text{Rx,a}}), \quad (6)$$
$$E_{\text{Rx,b}}(t) = \hat{E}_{\text{Rx,b}}\cos(\omega_{\text{Rx,b}}t + \varphi_{\text{Rx,b}}).$$

This leads to an oscillating power at frequency $\omega_{\text{Rx,THz}} = |\omega_{\text{Rx,a}} - \omega_{\text{Rx,b}}|$ with a phase $\varphi_{\text{Rx,THz}}$,

$$P_{\text{Rx,1}}(t) = \hat{P}_{\text{Rx,1}}\cos(\omega_{\text{Rx,THz}}t + \varphi_{\text{Rx,THz}}), \quad (7)$$

with the abbreviations

$$\hat{P}_{\text{Rx,1}} = \hat{E}_{\text{Rx,a}}\hat{E}_{\text{Rx,b}}, \quad \varphi_{\text{Rx,THz}} = \varphi_{\text{Rx,a}} - \varphi_{\text{Rx,b}}. \quad (8)$$

At the same time, the PIPED is biased with a DC voltage $U_{\text{Rx,0}}$, which is superimposed by the time-variant THz signal $U_{\text{Rx,1}}(t)$ generated by the THz antenna. The overall time-dependent voltage applied to the PIPED hence reads

$$U_{\text{Rx}}(t) = U_{\text{Rx,0}} + U_{\text{Rx,1}}(t), \quad (9)$$

where

$$U_{\text{Rx,1}}(t) = \hat{U}_{\text{Rx,1}}(t)\cos(\omega_{\text{Tx,THz}}t + \varphi_{\text{Tx,THz}}(t) - \varphi_{\text{TxRx}}). \quad (10)$$

In this relation, the phase at the receiver depends on the phase delay $\varphi_{\text{TxRx}}$ that the THz wave experiences when propagating from the Tx to the Rx. Due to the voltage-dependent PIPED sensitivity, the time-varying voltage $U_{\text{Rx}}(t)$ leads to a temporal variation of the sensitivity $S_{\text{Rx}}(U_{\text{Rx}}(t))$, oscillating at the frequency $\omega_{\text{Tx,THz}}$ of the incident THz wave. The PIPED photocurrent is given by the product of the time-variant sensitivity with the time-variant optical power $P_{\text{Rx,1}}(t)$. For the case of homodyne detection, $\omega_{\text{Rx,THz}} = \omega_{\text{Tx,THz}} = \omega_{\text{THz}}$, the baseband current at the output of the Rx PIPED is given by

$$I_{\text{BB}}(t) = I_{\text{BB,0}} + I_{\text{BB,1}}(t)$$
$$= I_{\text{BB,0}} + \hat{I}_{\text{BB,1}}(t)\cos(\varphi_{\text{BB}}(t)), \quad (11)$$

where the amplitude $\hat{I}_{\text{BB,1}}(t)$ of the time-variant part of the baseband photocurrent and the associated time-variant phase $\varphi_{\text{BB}}(t)$ are connected to the amplitude and the phase of the time-variant THz signal $U_{\text{Rx,1}}(t)$,

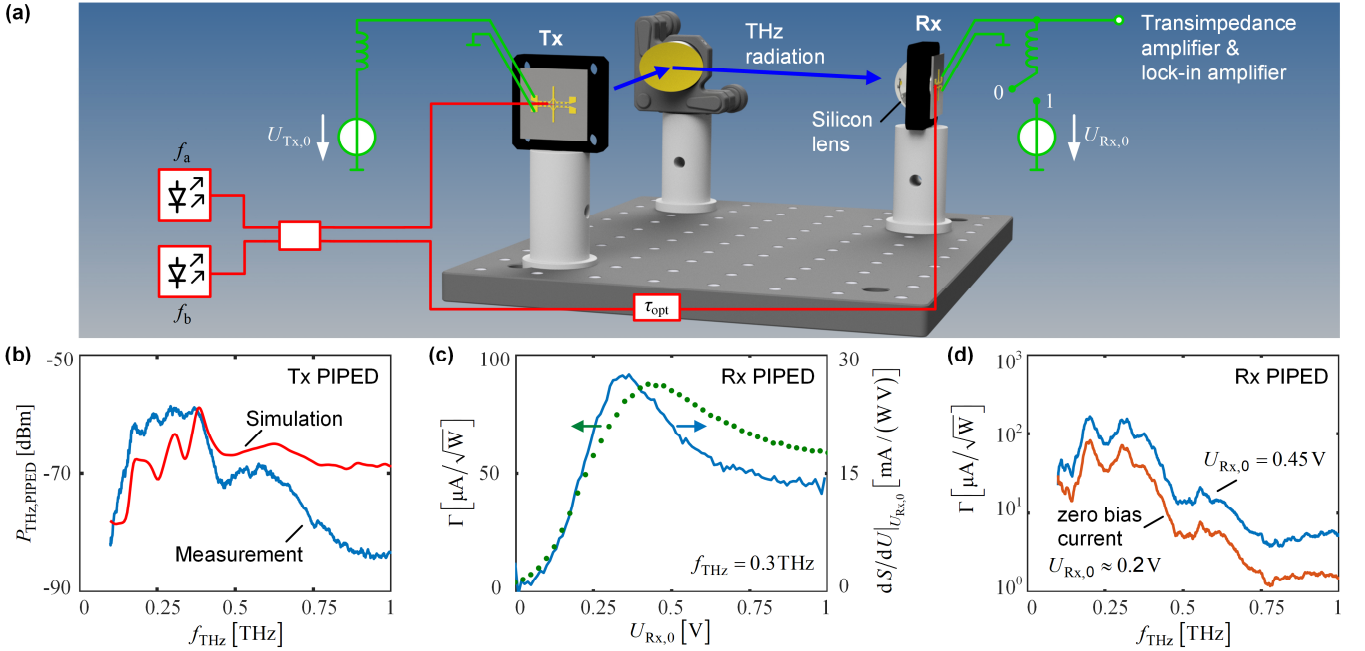

**Figure 3** Experimental demonstration of PIPED performance for O/T and T/E conversion **(a)** Experimental setup with free-space arrangement of Tx and Rx. For measuring the PIPED Tx (Rx) separately, the respective other component is replaced by commercially device in a free-space setup. Light from two lasers with frequencies $f_a$ and $f_b$ is superimposed to generate an optical power beat at $f_{THz} = |f_a - f_b|$, which simultaneously generates a THz wave in the PIPED Tx and acts as a phase-synchronous local oscillator (LO) for homodyne detection in the PIPED Rx. When using the PIPED at the Tx, the device is biased by a DC voltage $U_{Tx,0}$, which is modulated by an AC signal for lock-in detection. The bias is applied to the device via dedicated on-chip low-pass structures for decoupling of the THz component, see Inset 1 of Fig. 1. The same structures are used at the PIPED Rx, along with an additional bias-T in the feed circuit that allows to separate the DC bias from the AC signal. The AC signal is amplified by a transimpedance amplifier and detected by the lock-in amplifier. When operating the PIPED as a THz receiver, we either apply a defined bias voltage $U_{Rx,0}$ by connecting an external voltage source (switch position 1), or we leave the bias contacts open (switch position 0, "zero bias current"), which leads to a build-up of an internal forward-bias of $U_{Rx,0} \approx 0.2\,\text{V}$ upon illumination of the PIPED. **(b)** PIPED as a THz Tx. Measured (blue) and simulated radiated power (red). Details on the simulation can be found in the Methods. **(c)** PIPED as a THz Rx. Measured conversion factor $\Gamma$ (green) as a function of the Rx bias voltage $U_{Rx,0}$. The bias-dependent slope sensitivity $dS/dU|_{U_{Rx,0}}$ is also depicted (blue). Its maximum deviates slightly from the maximum of $\Gamma$ at the carrier frequency $0.3\,\text{THz}$, which we attribute to the voltage-dependent transit time. **(d)** Conversion factor $\Gamma = \hat{I}_{BB,1}/\sqrt{P_{THz}}$ of the Rx as a function of the THz frequency both for an externally applied DC bias of $U_{Rx,0} = 0.45\,\text{V}$ and for zero bias-current operation leading to an internal bias voltage of $U_{Rx,0} \approx 0.2\,\text{V}$. The maximum conversion factor is achieved for a bias voltage of $U_{Rx,0} = 0.45\,\text{V}$, see Subfigure (c).

$$\hat{I}_{BB,1}(t) = \frac{1}{2} \left.\frac{dS_{Rx}}{dU_{Rx}}\right|_{U_{Rx,0}} \hat{P}_{Rx,1} \hat{U}_{Rx,1}(t), \quad (12)$$

$$\varphi_{BB}(t) = \varphi_{Tx,THz}(t) - \varphi_{Rx,THz} + \varphi_{TxRx}.$$

For sensitive detection, the slope $dS_{Rx}/dU_{Rx}|_{U_{Rx,0}}$ of the sensitivity as a function of voltage has to be maximized, such that small variations of the THz voltage translate into large variations of the baseband photocurrent amplitude. The phase $\varphi_{BB}(t)$ may be properly adjusted with the variable time delay $\tau_{opt}$, see Fig. 3(a). A more convenient alternative measurement technique is explained in the Methods Section.

**Demonstration of T-wave generation and detection**
For an experimental demonstration of the PIPED performance in O/T and T/E conversion, we first characterize the Tx and the Rx separately. To this end, we fabricate a PIPED that is connected to an on-chip dipole antenna as shown in Inset 1 of Fig. 1. To supply a bias voltage $U_{Tx,0}$ to the PIPED, we use bias lines equipped with terahertz chokes that prevent leakage of THz signals from the antenna. With this configuration, the T-wave can be transmitted into and received from free space. For measuring the device performance, we use the setup depicted in Fig. 3(a), where the Tx and the Rx are driven by the same lasers for homodyne detection. To increase the sensitivity of T-wave detection, we use a modulated bias voltage $U_{Tx,0}$ that leads to a modulated THz power and helps in detecting the received T-wave with a lock-in amplifier. The THz wave is transmitted via a silicon lens and redirected to the lensed Rx antenna by an off-axis parabolic mirror.

For characterizing the PIPED Tx performance, a commercially available photoconductive THz receiver (Toptica, EK - 000725) is used. The THz power $P_{THz,PIPED}$ generated by the PIPED Tx leads to a current $I_{Rx,PIPED}$ in the reference receiver. For comparison, we measured the receiver current $I_{Rx,ref}$ when using a commercially available Tx (Toptica, EK - 000724) with a known THz power $P_{THz,ref}$. The PIPED THz output power can be estimated by

$$P_{\text{THz,PIPED}} = P_{\text{THz, ref}} \left( I_{\text{Rx,PIPED}} / I_{\text{Rx,ref}} \right)^2 . \quad (13)$$

The THz power $P_{\text{THz,PIPED}}$ generated by the PIPED is shown in Fig. 3(b). The PIPED is capable of generating radiation at frequencies of up to 1 THz. The measurements agree quite well with the simulated THz power (red), see Methods for details. The peaks in the simulation and the measurement are caused by antenna and bias line resonances. The roll-off at larger frequencies is in accordance with the bandwidth limitations expected due to the carrier transit time in the 100 nm wide silicon core of the PIPED. This width can be further reduced[29] e. g., to 75 nm, which decreases the transit time accordingly. The PIPED radiates a power of -59 dBm at 0.3 THz, limited by the maximum permissible photocurrent $I_{\text{Tx}}(t)$ in the 2 μm long detector. This limitation can be overcome by exploiting the small PIPED footprint, allowing to drive many PIPED in parallel, which may, e.g., feed the same multi-port antenna[42] and thereby dramatically increase the THz output power.

For evaluating the PIPED Rx performance, we use a commercially available Tx (Toptica, EK - 000724). According to Eq. (12), the baseband photocurrent amplitude $\hat{I}_{\text{BB,1}}$ depends linearly on the amplitude $\hat{U}_{\text{Rx,1}}$ of the THz voltage, which is proportional to the square root of the THz power $P_{\text{THz}}$. As a metric for the Rx sensitivity, we can hence define the ratio

$$\Gamma = \frac{\hat{I}_{\text{BB,1}}}{\sqrt{P_{\text{THz}}}} \propto \left. \frac{dS_{\text{Rx}}}{dU_{\text{Rx}}} \right|_{U_{\text{Rx,0}}} \hat{P}_{\text{Rx,1}} , \quad (14)$$

which describes the conversion factor from THz signal to baseband photocurrent. The linear relationship between $\hat{I}_{\text{BB,1}}$ and $\sqrt{P_{\text{THz}}}$ is experimentally confirmed, see Supplementary Section 2. The conversion factor depends on the sensitivity slope $\left. dS_{\text{Rx}}/dU_{\text{Rx}} \right|_{U_{\text{Rx,0}}}$, see Eq. (12). This is experimentally confirmed by measuring the conversion factor $\Gamma$ in dependence of the bias voltage $U_{\text{Rx,0}}$ at a frequency of 0.3 THz. The result is depicted in Fig. 3 (c) (green, left axis) along with the slope of the sensitivity (blue, right axis) derived from the static $S(U)$ characteristic in Fig. 2(c). The two curves are in fair agreement. We further verify that the conversion factor depends linearly on the incident optical power $\hat{P}_{\text{Rx,1}}$, see Supplementary Section 2. These findings confirm the validity of our PIPED model used to describe the Rx.

Similarly, we demonstrate the ability of a PIPED to perform broadband T/E conversion at the Rx. To this end, we measure the conversion factor $\Gamma$ as a function of the THz frequency for two cases. In a first measurement, we leave the bias contacts open, which is denoted as switch position 0 in Fig. 3(a) ("zero bias-current operation"). This leads to the build-up of an internal forward-bias of $U_{\text{Rx,0}} \approx 0.2\,\text{V}$ when illuminating the PIPED, see Supplementary Section 3. For the second measurement, we turn the switch in Fig. 3(a) to position 1 and connect an external DC voltage source $U_{\text{Rx,0}} = 0.45\,\text{V}$. The results of the measured conversion efficiencies $\Gamma$ are shown in Fig. 3(d). The PIPED is able to receive radiation at frequencies up to 1 THz and beyond. At a frequency of 0.3 THz, the devices exhibit a conversion factor of $150\,\mu\text{A}/\sqrt{\text{W}}$ for bias voltages of $U_{\text{Rx,0}} = 0.45\,\text{V}$ and of $65\,\mu\text{A}/\sqrt{\text{W}}$ for zero bias current, i.e., internal bias voltages of $U_{\text{Rx,0}} \approx 0.2\,\text{V}$. As expected from the results shown in Fig. 3 (c), the bias voltage of 0.45 V leads to a larger slope $\left. dS_{\text{Rx}}/dU_{\text{Rx}} \right|_{U_{\text{Rx,0}}}$ of the sensitivity and hence to a higher conversion factor as compared to the 0.2 V bias. The resonances in Fig. 3 (d) and the drop of the conversion factor for larger frequencies are caused by the frequency response of the antenna, the PIPED and the bias lines. In addition, we measured the short-circuit noise current $\sqrt{\langle I_{\text{N}}^2 \rangle}$ in a signal bandwidth $B$ for the case that there is no input T-wave signal[43]. For zero bias current, the device is driven by its internal bias voltage of 0.2 V only, and the noise current amounts to $12\,\text{pA}/\sqrt{\text{Hz}}$, which is similar to the to $15\,\text{pA}/\sqrt{\text{Hz}}$ reported for state-of-the-art photoconductors[13] and compares favorably with THz receivers based on photodiodes[44], for which an RMS current of $28\,\text{nA}/\sqrt{\text{Hz}}$ can be estimated from the noise floor of $-104.1\,\text{dBm/Hz}$ assuming a 50 Ω load. For better comparability to widely used THz spectroscopy systems, we also determine the signal-to-noise power ratio (SNR) obtained in our experiment. Using the Toptica reference transmitter (EK-000724), the T-wave power incident on the Rx at 0.3 THz amounts to $P_{\text{THz}}$ = -23 dBm and leads to an SNR of 82 dB Hz$^{-1}$. This is comparable to values of 91 dB Hz$^{-1}$ (ref. 13) and 57 dB Hz$^{-1}$ (ref. 45) which have been achieved with conventional III-V devices, see Methods Section. For the externally applied bias voltage $U_{\text{Rx,0}} = 0.45\,\text{V}$, the RMS noise current increases considerably to values of $270\,\text{pA}/\sqrt{\text{Hz}}$, thus reducing the achieved SNR at 0.3 THz to 62 dB Hz$^{-1}$ despite the larger conversion factor. The strong noise level for the case of an externally applied bias voltage is attributed to the DC voltage source. Improved noise levels might be achieved by low-pass filtering the bias voltage.

**Monolithically integrated T-wave system**

To demonstrate the technological advantages of the PIPED concept, we monolithically integrate arrays of PIPED Tx and PIPED Rx on a common silicon chip. Tx and Rx are coupled by short T-wave transmission lines having various lengths $L$ between 10 μm and approximately 1 mm. An SEM picture of such a Tx-Rx pair is displayed in Fig. 4 (a). A 1 μm-wide gap in the middle of the T-wave transmission line acts as a DC block to decouple the bias voltages of the Tx and the Rx. The insets show magnified pictures of the Tx, Rx and the gap. In the experiment, we use the PIPED Tx and Rx to measure the amplitude and phase transfer characteristics of the transmission line at THz frequencies. Optical signals are fed to the Tx and the Rx PIPED by separate silicon photonic waveguides equipped with grating couplers. The setup for measuring the transmission line characteristics is shown in Fig. 4 (b). As in the previous experiments, the Tx and Rx are fed with two optical spectral lines $f_a$ and $f_b$ at varying frequency separations. The Tx is biased in forward direction $U_{Tx,0} > 0$ and feeds the T-wave transmission line. At the end of the transmission line, the Rx PIPED acts as a homodyne receiver to coherently detect the T-wave signal and down-converts it to a baseband current $I_{BB,1}$, see Eqs. (11) and (12). The Rx PIPED is operated under zero bias-current condition to minimize noise. Because the PIPED features a large impedance at THz frequencies, all transmission lines are essentially terminated by open circuits at both ends and thus act as T-wave resonators, for which the resonance frequencies are dictated by the respective geometrical length $L$. At the Rx, we measure the baseband photocurrent $I_{BB,1}$. By sweeping the Tx frequency, we can separate the amplitude $\hat{I}_{BB,1}$ from the phase $\varphi_{BB}$, see Methods for details. To quantify the transmission line transfer characteristics, we use the complex transfer impedance $\underline{Z}_{21} = |\underline{Z}_{21}| e^{j\varphi_{21}}$, which is defined as the ratio of the complex T-wave voltage amplitude at the Rx PIPED and the complex T-wave current amplitude at the Tx PIPED. The magnitude of the complex transfer impedance is directly proportional to the baseband current amplitude $\hat{I}_{BB,1}$,

$$|\underline{Z}_{21}| = a\hat{I}_{BB,1}, \qquad (15)$$

where the proportionality factor $a$ depends on the operation conditions of the Tx and the Rx PIPED,

$$a = \left(\frac{1}{2}\frac{dS_{Rx}}{dU_{Rx}}\bigg|_{U_{Rx,0}} \hat{P}_{Rx,1} S_{Tx}(U_{Tx,0}) \hat{P}_{Tx,1}\right)^{-1}, \qquad (16)$$

see Supplementary Section 4. The corresponding phase $\varphi_{21}$ is obtained from $\varphi_{BB}$ by swapping the role of the Tx and the Rx PIPED, thus allowing to eliminate the influence of the unknown group delay in the optical fibers, see Methods. The measured magnitude and phase characteristics of the transfer impedance $\underline{Z}_{21,meas}$ are depicted as blue traces in Fig. 4(c) - (d). For comparison, we have also numerically calculated the transfer impedance of the T-wave transmission lines using a commercially available time-domain solver[46]. The results of the simulated transfer impedance $\underline{Z}_{21,\text{sim}}$ are indicated as red traces in Fig 4(c) - (d). Note that calculating the proportionality factor $a$ according to Eq. (16) is subject to large uncertainties since the experimental operating conditions of the Tx and Rx PIPED at THz frequencies are only approximately known. For estimating $|\underline{Z}_{21}|$ from the measured baseband photocurrent amplitude $\hat{I}_{\text{BB},1}$, we have therefore chosen $a$ such that we obtain best agreement of simulation and measurement, see Section 5 of the Supplementary Information. Moreover, we numerically correct for the frequency roll-off of the PIPED when deriving the measured transfer impedance $\underline{Z}_{21,\text{meas}}$ from the baseband photocurrent amplitude $\hat{I}_{\text{BB},1}$. To this end, the transit-time limited frequency response of both PIPED was approximated by an RC lowpass characteristic with a corner frequency of $0.3\,\text{THz}$, obtained from a least-squares fit of the measurement data, see Supplementary Section 5. The measured and the simulated transmission line characteristics show perfect agreement with respect to amplitude and phase over the entire frequency range $f_{\text{THz}} = (0.05\ldots 0.8)\,\text{THz}$, corresponding to a span of 1.5 decades. The field distributions along the transmission lines have been investigated by numerical simulations of the structures, see Supplementary Section 6. These findings show that signal processing in PIPED can be quantitatively described by a reliable mathematical model, thus enabling deterministic photonic-electronic signal processing over an ultra-broadband frequency range.

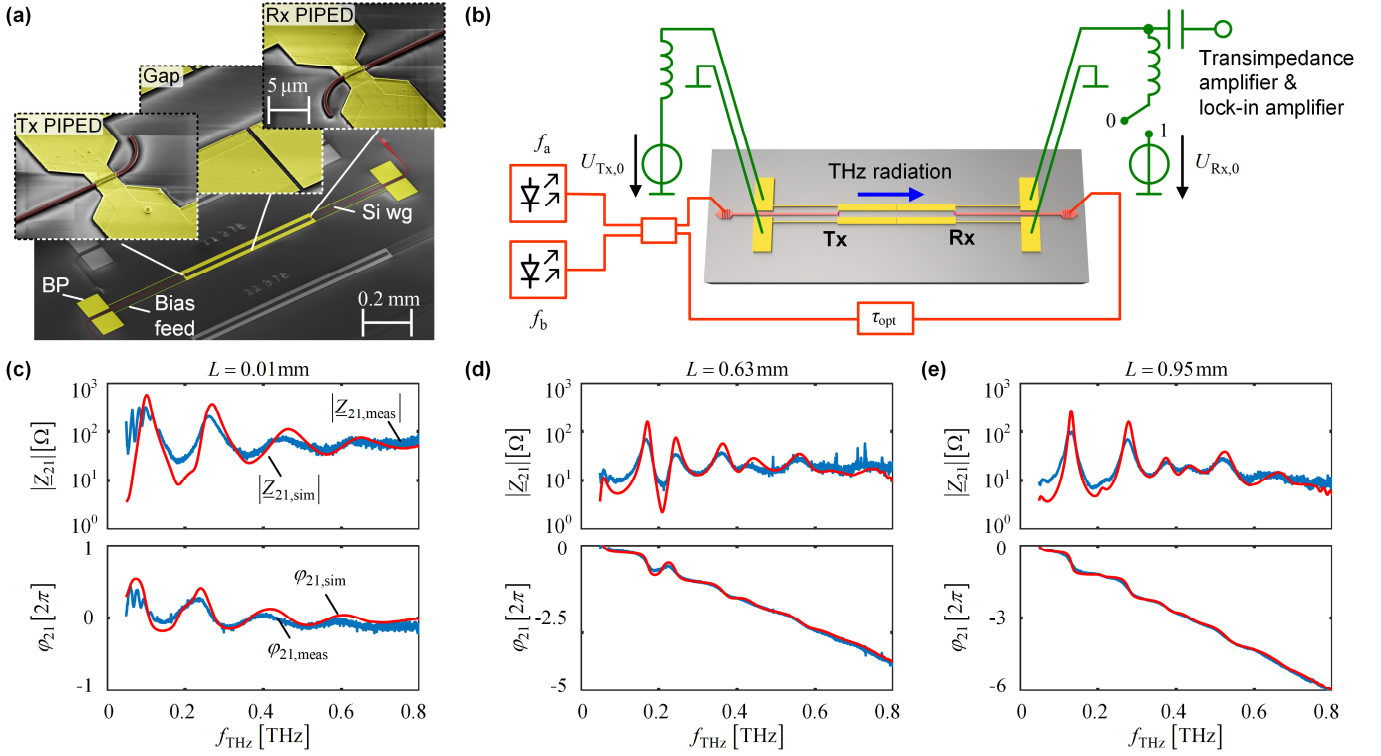

**Figure 4** Photonic THz transmitter and homodyne THz receiver on a silicon chip. (a) SEM picture of the THz system consisting of a THz Tx, a coplanar transmission line, and a THz Rx. Light is coupled to the Tx and Rx by silicon waveguides. Bias pads (BP) and THz-blocking bias feeds allow to apply a voltage, or to extract a signal current to or from the PIPED, respectively. A gap in the middle of the transmission line separates the bias voltages of Tx and the Rx. The inset shows magnified pictures of Tx, Rx and transmission line gap. (b) Experimental setup for the on-chip transmission line measurements. The Tx and Rx are fed with two optical spectral lines $f_a$ and $f_b$ at varying frequency separations. The Tx is biased in forward direction $U_{Tx,0} > 0$ and feeds the T-wave transmission line. At the end of the transmission line, the Rx PIPED acts as a homodyne receiver to coherently detect the T-wave signal by down-conversion to the baseband current $I_{BB,1}$. To reduce the noise, the switch is set to 0 to operate the Rx PIPED with zero-bias current. (c)-(e) Magnitude and phase characteristics of the transfer impedance $\underline{Z}_{21} = |\underline{Z}_{21}| \exp(j\varphi_{21})$. The red curves shows the simulated curves, $|\underline{Z}_{21,sim}|$ and $\varphi_{21,sim}$, for three transmission line lengths $L = (0.01, 0.63, 0.95)\,\text{mm}$ and the blue curves shows the corresponding measurement results, $|\underline{Z}_{21,meas}|$ and $\varphi_{21,meas}$. To account for the frequency roll-off of the PIPED, we use an RC-low pass characteristic with a 3dB-frequency of $f_{3dB} = 0.3\,\text{THz}$, see Supplementary Section 5. The measurements (blue) agree very well with the simulation (red).

**Summary and outlook**

In summary, we have demonstrated a novel approach to T-wave signal processing that exploits internal photoemission at metal-semiconductor interfaces of silicon-plasmonic structures. We show that T-wave transmitter and receiver can be monolithically co-integrated on the silicon photonic platform, thus enabling a novel class of photonic-electronic signal processors that may exploit the outstanding technical maturity and performance of the silicon photonic device portfolio. When used as a T-wave Rx, the performance of our current PIPED devices can already compete with state-of-the-art III-V photoconductors, while further improvements are still possible by optimized device geometry and materials. The devices are analyzed and described by a quantitatively reliable mathematical model. In a proof-of-concept experiment, we use co-integrated PIPED transmitters and receivers for measuring the complex transfer function of an integrated T-wave transmission line.

# Methods

**PIPED device fabrication.** The PIPED are fabricated on commercially available silicon-on-insulator (SOI) substrates with a 340 nm thick device layer. The waveguide structures are defined by electron beam lithography and reactive ion etching. Directional metal evaporation under various angles allows to deposit different metals on each side of the silicon core and enables the fabrication of an asymmetrically metallized silicon core which has a width as small as 100 nm and below[29]. To keep the fabrication simple, we use gold (Au) and titanium (Ti) to metallize the silicon core sidewalls since these materials were readily available in our laboratory. Note that other metals such as copper or aluminum with similar optical and electrical properties could be used as well, thus allowing PIPED fabrication using standard CMOS materials. During metallization, an undercut hard mask on top of the silicon core is used to prevent a short circuit between the Au and the Ti region, see Fig. 2(a). For more information on the fabrication of the PIPED see ref. 29.

**Coherent T-wave detection.** For sensitive detection of the Rx current $\hat{I}_{\text{BB},1}(t)\cos(\varphi_{\text{BB}})$, a lock-in amplifier is used. To this end, we modulate the Tx bias voltage $U_{\text{Tx},0}$, which leads to a modulation of the THz Tx current $\hat{I}_{\text{Tx},1}(t)$, of the received THz voltage $\hat{U}_{\text{Rx},1}(t)$, and of the time-variant part $\hat{I}_{\text{BB},1}$ of the baseband photocurrent, see Eq. (12). The modulation frequency is set to $f_{\text{LI}} = 12.2\,\text{kHz}$, and the modulated receiver current $\hat{I}_{\text{BB},1}(t)\cos(\varphi_{\text{BB}})$ is measured by a lock-in amplifier (Toptica Sys DC 110). The phase $\varphi_{\text{BB}}$ of the baseband current depends on the phase $\varphi_{\text{Tx,THz}}$ of the optical power beat at the Tx, the phase $\varphi_{\text{Rx,THz}}$ of the optical power beat at the Rx, as well as on the phase delay $\varphi_{\text{TxRx}}$ gathered during transmission of the THz wave from the Tx to the Rx,

$$\varphi_{\text{BB}} = \varphi_{\text{Tx,THz}} - \varphi_{\text{Rx,THz}} + \varphi_{\text{TxRx}} \qquad (17)$$

For homodyne detection, see Fig. 3(a) and Fig. 4(b), the phase difference $\varphi_{\text{Tx,THz}} - \varphi_{\text{Rx,THz}}$ of the optical power beats depends only on the optical group delay $\tau_{\text{opt}}$ between the Tx and the Rx,

$$\varphi_{\text{Tx,THz}} - \varphi_{\text{Rx,THz}} = \omega_{\text{THz}}\tau_{\text{opt}} = 2\pi f_{\text{THz}} n_{\text{g}}\,\Delta L/c \qquad (18)$$

where $c$ is the speed of light, $n_{\text{g}}$ the optical group refractive index of the single-mode fibers and $\Delta L$ the path length difference of the optical wave. In this relation, we assumed that the on-chip SOI waveguide at the Tx and the Rx have the same length. Scanning the frequency $f_{\text{THz}}$ leads to an oscillation of the current $\hat{I}_{\text{BB},1}\cos(\varphi_{\text{BB}})$, where the number of oscillations per frequency interval depends on the time delay $\tau_{\text{opt}}$. We record the current $\hat{I}_{\text{BB},1}\cos(\varphi_{\text{BB}})$ as a function of the THz frequency $f_{\text{THz}}$ and separate the amplitude $\hat{I}_{\text{BB},1}$ from the phase $\varphi_{\text{BB}}$ by means of a Fourier transformation of the received signal $\hat{I}_{\text{BB},1}\cos(\varphi_{\text{BB}}) = \tfrac{1}{2}\hat{I}_{\text{BB},1}\left[e^{j\varphi_{\text{BB}}} + e^{-j\varphi_{\text{BB}}}\right]$. After Fourier transformation, we numerically remove the negative frequency components and calculate the inverse Fourier transform to obtain the complex amplitudes $\tfrac{1}{2}\hat{I}_{\text{BB},1}e^{j\varphi_{\text{BB}}}$ of the frequency-dependent baseband current. From these complex amplitudes, we can directly read the received current amplitude $\hat{I}_{\text{BB},1}$ and the phase $\varphi_{\text{BB}}$. Regarding the phase information, we are mainly interested in the phase delay $\varphi_{\text{TxRx}}$ accumulated by the THz wave during propagation from the Tx to the Rx. To obtain this information, we need to eliminate the phase shifts that are caused by the unknown optical group delay $\tau_{\text{opt}}$ between the Tx and the Rx. To this end, we perform two consecutive measurements and exploit the fact that the same PIPED can act both as Tx and as Rx. The first measurement is performed as described in Fig 4(b). In the second measurement we swap the role of Tx and Rx by switching the bias conditions of the associated PIPED. Since the optical beam path is unchanged, only the sign of $\varphi_{\text{Tx,THz}} - \varphi_{\text{Rx,THz}}$ has changed, while $\varphi_{\text{TxRx}}$ is unchanged. For Eq. (17), the THz phase can then be calculated,

$$\varphi_{\text{TxRx}} = (\varphi_{\text{BB,norm}} + \varphi_{\text{BB,rev}})/2 \qquad (19)$$

The amplitude of the baseband current $\hat{I}_{\text{BB},1,\text{rev}}$ is not changed by reversing Tx and Rx. In Fig. 4(c)-(e) the plotted baseband current $\hat{I}_{\text{BB},1}$ was taken as the mean of both measurements

$$\hat{I}_{\text{BB},1} = (\hat{I}_{\text{BB},1,\text{norm}} + \hat{I}_{\text{BB},1,\text{rev}})/2 \qquad (20)$$

**3D EM simulations.** The response of the THz dipole antenna, see Fig. 3(b), and the THz resonators, see Fig. 4(c)-(e), are simulated using a commercially available numerical time domain solver (CST Microwave Studio[46]). A crucial aspect is the correct representation of the thin metal layers of the transmission lines and the antennae. Atomic force microscope measurements reveal a 110 nm-thick gold layer on top of a 40 nm-thick titanium layer. In the simulation of transmission lines with dimensions in the mm-range, a detailed representation of the field within the metal would lead to an unrealistic number of mesh cells that cannot be handled with the available computing resources. Therefore, we use the "thin panel" option of CST microwave studio, which relates the tangential electric and magnetic fields on the surface of the metal with the help of surface impedances[46]. The Rx and the Tx PIPED are described by discrete equivalent circuits. For simulation of the Tx as depicted in Fig. 3(b), the PIPED is represented by an ideal current source with an amplitude of 50 µA that excites the Tx antenna. The current amplitude of 50 µA was estimated from the DC current generated in the PIPED when using the respective optical input power. The antenna is simulated as a thin metal panel on a silicon substrate with open boundaries to emulate an infinite half-space below. This approach is valid because the silicon chip is in direct contact with a macroscopic silicon lens with dimensions much larger than the THz wavelength or the Tx antenna. The total radiated power corresponds to the total power radiated in the lower silicon half-space. For a numerical analysis of the THz resonator in Fig. 4, all transmission line structures are again represented as thin metal panels deposited on a semi-infinite silicon substrate. The Tx PIPED is modeled as a current source, and the Rx PIPED is modeled as an open circuit across which the voltage is measured. In both cases, we obtain essentially the same results when replacing the ideal current source or open circuit with infinite internal resistances by devices that feature finite internal resistances of 10 kΩ. This is a worst-case estimate of the real internal PIPED resistance which is typically well above the assumed value of 10 kΩ, see Supplementary Section 4. The complex transfer impedance can be calculated by dividing the simulated THz voltage amplitude at the Rx by the THz current amplitude at the Tx. Alternatively, the complex transfer impedance $\underline{Z}_{21}$ can be derived from numerically calculated S-parameters of the THz resonator, see Supplementary Section 7. Both methods lead to the same results.

**Optical setup.** For generating the optical power beat two temperature-controlled tunable distributed feedback (DFB) lasers with linewidths of around 1 MHz are used (#LD-1550-0040-DFB-1). The difference frequency can be tuned from 0 to 1.2 THz with a relative frequency accuracy better than 10 MHz and an absolute frequency accuracy of 2 GHz. For feed of light from single-mode fibers to antenna-coupled PIPED, Fig. 3(a), we use grating couplers with losses of 5 dB, followed by 0.9 mm-long on-chip SOI waveguides with propagation losses of 2.3 dB/mm, thus leading to total coupling loss of 7 dB. For the integrated THz system, Fig. 4(b), the grating coupler losses amount to 5.3 dB, the waveguide propagation losses to 1.1 dB/mm, and the length of the on-chip waveguide is 0.6 mm. This leads to total coupling losses of 6 dB. The optical path length difference $\Delta L$ between the Tx and the Rx was between 1.0 m and 1.3 m for the antenna-coupled PIPED, Fig. 3(a), and 0.6 m for the integrated THz system, Fig. 4. For the antenna-coupled PIPED, the THz Tx and the Rx characteristics were sampled with steps of 10 MHz, Fig. 3(b) and (d), and the integration time of the lock-in amplifier was chosen between 20 ms and 100 ms. For spectral characterization of the integrated THz systems, Fig. 4(c)-(d), sampling points were taken every 25 MHz, and the integration time was set to 100 ms.

**SNR estimation.** In the paper, we give the signal-to-noise (SNR) power ratio in units of dB Hz$^{-1}$, which is independent of the integration time of the lock-in detection. Other groups[13,45] specify the SNR in dB in combination with the integration time $T$ of the lock-in detector. To compare these values to our results, we convert the SNR of ref. 13 and ref. 45 to dB Hz$^{-1}$. From Fig.3 in ref. 13 we read an SNR of 93 dB at 300 GHz. In this measurement the lock-in filter slope was set to 12 dB/octave and an integration time of $T = 200$ ms was used. This leads to an equivalent noise bandwidth of $B = 1/(8T) = 0.625\,\text{Hz}$ and an SNR value of 91 dB Hz$^{-1}$ as stated in the main text. In ref. 45, the measured SNR value is 47 dB and an integration time of 23.4 ms was used. The filter slope is not explicitly stated. By assuming a filter slope of 6 dB/octave, we get an equivalent noise bandwidth of $B = 1/(4T) = 10.7\,\text{Hz}$ and therefore an SNR of 57 dB Hz$^{-1}$.

## Acknowledgements

This work was supported by the ERC Starting Grant 'EnTeraPIC', number 280145, by the Alfried Krupp von Bohlen und Halbach Foundation, by the Helmholtz International Research School of Teratronics (HIRST), by the Karlsruhe School of Optics and Photonics (KSOP), by Karlsruhe Nano Micro Facility (KNMF).

# Silicon-plasmonic integrated circuits for terahertz signal generation and coherent detection
## Supplementary Information

T. Harter[1,2*], S. Muehlbrandt[1,2], S. Ummethala[1,2], A. Schmid[1], L. Hahn[2], W. Freude[1], C. Koos[1,2**]

[1]*Institute of Photonics and Quantum Electronics (IPQ), Karlsruhe Institute of Technology (KIT), 76131 Karlsruhe, Germany*
[2]*Institute of Microstructure Technology (IMT), Karlsruhe Institute of Technology (KIT), 76131 Karlsruhe, Germany*
[*]tobias.harter@kit.edu, [**]christian.koos@kit.edu

## 1. Detailed derivation of formulae

The plasmonic internal photoemission detector (PIPED) can be used as a T-wave transmitter (Tx) as well as a T-wave receiver (Rx). In the following, a detailed explanation of both functionalities is given. As a Tx, the PIPED multiplies two time-dependent optical signals $E_{\text{Tx,a}}(t)$ and $E_{\text{Tx,b}}(t)$ to produce a photocurrent that corresponds to the difference-frequency waveform (O/T conversion). In the following, we assume that the optical signal $E_{\text{Tx,a}}(t)$ oscillates at frequency $f_{\text{Tx,a}} = \omega_{\text{Tx,a}}/(2\pi)$ and carries a (normalized) amplitude modulation $\hat{E}_{\text{Tx,a}}(t)$ and/or a phase modulation $\varphi_{\text{Tx,a}}(t)$, whereas the optical signal $E_{\text{Tx,b}}(t)$ is simply a CW carrier with constant amplitude $\hat{E}_{\text{Tx,b}}$, frequency $f_{\text{Tx,b}} = \omega_{\text{Tx,b}}/(2\pi)$, and phase $\varphi_{\text{Tx,b}}$,

$$E_{\text{Tx,a}}(t) = \hat{E}_{\text{Tx,a}}(t)\cos(\omega_{\text{Tx,a}}t + \varphi_{\text{Tx,a}}(t)),$$
$$E_{\text{Tx,b}}(t) = \hat{E}_{\text{Tx,b}}\cos(\omega_{\text{Tx,b}}t + \varphi_{\text{Tx,b}}). \quad (S1)$$

The superposition of the two optical signals leads to an optical power $P_{\text{Tx}}(t)$, which oscillates with the difference angular THz frequency $\omega_{\text{Tx,THz}} = |\omega_{\text{Tx,a}} - \omega_{\text{Tx,b}}|$ and shows a phase $\varphi_{\text{Tx,THz}}(t) = \varphi_{\text{Tx,a}}(t) - \varphi_{\text{Tx,b}}$,

$$\begin{aligned}P_{\text{Tx}}(t) &= \left\langle \left(E_{\text{Tx,a}}(t) + E_{\text{Tx,b}}(t)\right)^2 \right\rangle_{1/f_{\text{Tx,b}}} \\ &= \tfrac{1}{2}\left(\hat{E}_{\text{Tx,a}}^2(t) + \hat{E}_{\text{Tx,b}}^2\right) \\ &\quad + \hat{E}_{\text{Tx,a}}(t)\hat{E}_{\text{Tx,b}}\cos(\omega_{\text{Tx,THz}}t + \varphi_{\text{Tx,THz}}(t)) \\ &= P_{\text{Tx,0}}(t) + \underbrace{\hat{P}_{\text{Tx,1}}(t)\cos(\omega_{\text{Tx,THz}}t + \varphi_{\text{Tx,THz}}(t))}_{P_{\text{Tx,1}}(t)}\end{aligned} \quad (S2)$$

The average can be performed over either period of $E_{\text{Tx,a,b}}(t)$. This oscillating power is incident on the PIPED which is biased at a voltage $U_{\text{Tx,0}}$ and has a sensitivity $S_{\text{Tx}}(U_{\text{Tx,0}})$. The PIPED output current is in proportion to the optical power, $I_{\text{Tx}}(t) \sim P_{\text{Tx}}(t)$, and can be separated in a slowly varying term $I_{\text{Tx,0}}(t) \sim P_{\text{Tx,0}}(t)$ and a term $I_{\text{Tx,1}}(t) \sim P_{\text{Tx,1}}(t)$ which oscillates at the THz frequency,

$$\begin{aligned}I_{\text{Tx}}(t) &= S_{\text{Tx}}(U_{\text{Tx,0}})P_{\text{Tx}}(t) \\ &= I_{\text{Tx,0}}(t) + I_{\text{Tx,1}}(t) \\ &= I_{\text{Tx,0}}(t) + \hat{I}_{\text{Tx,1}}(t)\cos(\omega_{\text{Tx,THz}}t + \varphi_{\text{Tx,THz}}(t)).\end{aligned} \quad (S3)$$

In this relation we used the abbreviations

$$\begin{aligned}I_{\text{Tx,0}}(t) &= S_{\text{Tx}}(U_{\text{Tx,0}})\tfrac{1}{2}\left(\hat{E}_{\text{Tx,a}}^2(t) + \hat{E}_{\text{Tx,b}}^2\right), \\ \hat{I}_{\text{Tx,1}}(t) &= S_{\text{Tx}}(U_{\text{Tx,0}})\hat{E}_{\text{Tx,a}}(t)\hat{E}_{\text{Tx,b}}.\end{aligned} \quad (S4)$$

Hence, any modulation of the amplitude $\hat{E}_{\text{Tx,a}}(t)$ or the phase $\varphi_{\text{Tx,a}}(t)$ of the optical carrier translates directly to the THz range. The THz current is radiated by an antenna or coupled to a transmission line. Note that for lock-in detection, the transmitter DC bias $U_{\text{Tx,0}}$ could be slowly modulated, leading to a time-dependent quantity $U_{\text{Tx,0}}(t)$.

Similarly, the PIPED can be used for T-wave reception (T/E conversion). In this case, the device combines two functionalities, namely the generation of a THz local oscillator (LO) from two optical carriers, and the down-conversion of the received THz signal to the baseband. To this end, the PIPED is fed by a superposition of two unmodulated optical carriers, oscillating at angular frequencies $\omega_{\text{Rx,a}}$ and $\omega_{\text{Rx,b}}$ and having phases $\varphi_{\text{Rx,a}}$ and $\varphi_{\text{Rx,b}}$,

$$\begin{aligned}E_{\text{Rx,a}}(t) &= \hat{E}_{\text{Rx,a}}\cos(\omega_{\text{Rx,a}}t + \varphi_{\text{Rx,a}}), \\ E_{\text{Rx,b}}(t) &= \hat{E}_{\text{Rx,b}}\cos(\omega_{\text{Rx,b}}t + \varphi_{\text{Rx,b}}).\end{aligned} \quad (S5)$$

The superposition of the two optical signals in the Rx has an optical power $P_{\text{Rx}}(t)$, which oscillates with the difference angular frequency $\omega_{\text{Rx,THz}} = |\omega_{\text{Rx,a}} - \omega_{\text{Rx,b}}|$ and shows a phase $\varphi_{\text{Rx,THz}} = \varphi_{\text{Rx,a}} - \varphi_{\text{Rx,b}}$,

$$\begin{aligned}P_{\text{Rx}}(t) &= \left\langle \left(E_{\text{Rx,a}}(t) + E_{\text{Rx,b}}(t)\right)^2 \right\rangle_{1/f_{\text{Rx,b}}} \\ &= \tfrac{1}{2}\left(\hat{E}_{\text{Rx,a}}^2 + \hat{E}_{\text{Rx,b}}^2\right) \\ &\quad + \hat{E}_{\text{Rx,a}}\hat{E}_{\text{Rx,b}}\cos(\omega_{\text{Rx,THz}}t + \varphi_{\text{Rx,THz}}) \\ &= P_{\text{Rx,0}} + \underbrace{\hat{P}_{\text{Rx,1}}\cos(\omega_{\text{Rx,THz}}t + \varphi_{\text{Rx,THz}})}_{P_{\text{Rx,1}}(t)}.\end{aligned} \quad (S6)$$

At the same time, the PIPED is DC-biased with a voltage $U_{\text{Rx,0}}$ and the antenna superimposes a time-variant THz signal $U_{\text{Rx,1}}(t)$. The total time-dependent voltage applied to the PIPED contacts hence reads

$$U_{Rx}(t) = U_{Rx,0} + U_{Rx,1}(t), \tag{S7}$$

where we introduced the abbreviation

$$U_{Rx,1}(t) = \hat{U}_{Rx,1}(t)\cos(\omega_{Tx,THz}t + \varphi_{Tx,THz}(t) + \varphi_{TxRx}). \tag{S8}$$

In this relation, the phase at the receiver depends on the phase shift $\varphi_{TxRx}$ that the THz wave experiences when propagating from the Tx to the Rx. The voltage $U_{Rx}(t)$ leads to a time-variant PIPED sensitivity $S_{Rx}(U_{Rx}(t))$. The small-signal approximation reads

$$S_{Rx}(U_{Rx}(t)) = S_{Rx}(U_{Rx,0}) + \left.\frac{dS_{Rx}}{dU_{Rx}}\right|_{U_{Rx,0}} U_{Rx,1}(t). \tag{S9}$$

The PIPED photocurrent $I_{Rx}(t)$ is given by the product of this time-variant sensitivity $S_{Rx}(U_{Rx}(t))$ and the time-variant optical power $P_{Rx}(t)$,

$$I_{Rx}(t) = S_{Rx}(U_{Rx}(t))P_{Rx}(t). \tag{S10}$$

Averaging over a THz cycle leads to the down-converted baseband current

$$\begin{aligned}I_{BB}(t) &= \langle I_{Rx}(t)\rangle_{THz} \\ &= I_{BB,0} + I_{BB,1}(t) \\ &= I_{BB,0} + \hat{I}_{BB,1}(t)\cos((\omega_{Tx,THz} - \omega_{Rx,THz})t + \varphi_{BB}(t)),\end{aligned} \tag{S11}$$

where we used the abbreviations

$$I_{BB,0} = S_{Rx}(U_{Rx,0})P_{Rx,0},$$

$$\hat{I}_{BB,1}(t) = \frac{1}{2}\left.\frac{dS_{Rx}}{dU_{Rx}}\right|_{U_{Rx,0}} \hat{P}_{Rx,1}\hat{U}_{Rx,1}(t), \tag{S12}$$

$$\varphi_{BB}(t) = \varphi_{Tx,THz}(t) - \varphi_{Rx,THz} + \varphi_{TxRx}.$$

For the special case of homodyne reception, $\omega_{Rx,THz} = \omega_{Tx,THz} = \omega_{THz}$, the baseband photocurrent at the output of the Rx PIPED is given by

$$I_{BB}(t) = I_{BB,0} + \hat{I}_{BB,1}(t)\cos(\varphi_{BB}(t)). \tag{S13}$$

The time-variant part $I_{BB,1}(t)$ of the photocurrent is proportional to the voltage amplitude $\hat{U}_{Rx,1}(t)$ of Eq. (S8) and therefore in proportion to the incoming THz field.

## 2. PIPED used as T-wave Rx

In the following, we assume reception of an unmodulated THz voltage amplitude $\hat{U}_{Rx,1} \propto \sqrt{P_{THz}}$ corresponding to a received THz power $P_{THz}$, and we experimentally confirm the linear relationship between the baseband photocurrent amplitude $\hat{I}_{BB,1}$ and $\hat{U}_{Rx,1}$ as formulated in Eq. (S12).

We prove $\hat{I}_{BB,1} \propto \sqrt{P_{THz}}$ by measuring the baseband current $\hat{I}_{BB,1}$ for various THz powers $P_{THz}$ at a frequency of 0.19 THz, see Fig. S1. The slope defines the conversion factor

$$\Gamma = \frac{\hat{I}_{BB,1}}{\sqrt{P_{THz}}} \propto \left.\frac{dS_{Rx}}{dU_{Rx}}\right|_{U_{Rx,0}} \hat{P}_{Rx,1}. \tag{S14}$$

The data plotted in Fig. S1 correspond to a conversion factor of $110\,\mu A/\sqrt{W}$. In the main text we showed experimentally that the conversion factor $\Gamma$ is virtually proportional to the slope $\left.dS_{Rx}/dU_{Rx}\right|_{U_{Rx,0}}$ of the sensitivity, see Fig. 3(c). Additionally it is expected from Eq. (S14) that $\Gamma$ depends linearly on the optical power amplitude $\hat{P}_{Rx,1}$, which is experimentally confirmed by measurements shown in Fig. S2.

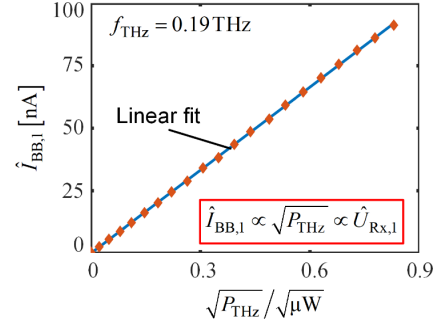

**Figure S1:** Baseband current amplitude $\hat{I}_{BB,1}$ in dependence of input THz power amplitude $P_{THz}$ at a frequency of 0.19 THz. The linear fit of the measurement demonstrates the excellent linearity between $\hat{I}_{BB,1}$ and $\sqrt{P_{THz}}$.

## 3. Operating conditions of Rx PIPED

Figure 3(a) and 4(b) of the main text show two operating conditions for the Rx PIPED. First, we connect an external DC voltage source $U_{Rx,0}$ by turning the switch to position 1. This allows to set the bias voltage to $U_{Rx,0} = 0.45\,V$ which maximizes the sensitivity slope $\left.dS_{Rx}/dU_{Rx}\right|_{U_{Rx,0}}$ and the receiver conversion factor $\Gamma$, see Fig. 3(c). Second, we turn the switch to position 0 (open circuit condition). In this case, no DC current flows. We call this situation "zero bias-current operation". The benefit of method 2 is that the effective noise current is reduced from $270\,pA/\sqrt{Hz}$ to $12\,pA/\sqrt{Hz}$. The large current noise for the case of an externally applied bias voltage is attributed to the DC voltage source. Shot noise is less important for bias currents in the order of $100\,\mu A$, which leads to an calculated effective shot noise current of only $5\,pA/\sqrt{Hz}$.

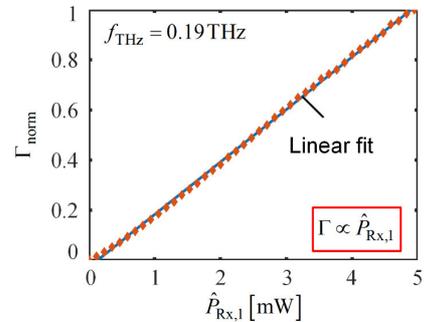

**Figure S2:** Normalized conversion factor $\Gamma_{norm}$ in dependence of the optical power amplitude $\hat{P}_{Rx,1}$, measured at a THz frequency of 0.19 THz. The conversion factor is normalized to the value obtained for the largest optical power amplitude.

If an Rx PIPED under zero-bias-current operation is illuminated with light, a voltage of roughly $U_{Rx,0} \approx 0.2\,\text{V}$ builds up, which acts as a forward bias for the PIPED. The build-up potential is estimated from the static $I_{Rx,0}$-$U_{Rx,0}$ (current-voltage) characteristic of the receiver PIPED illuminated with optical amplitudes $\hat{E}_{Rx,a}=0$, $\hat{E}_{Rx,b}=\text{const}$, see Fig. S3. The results are shown for three illumination conditions, $P_{Rx,0}=0\,\text{mW}$, $1\,\text{mW}$ and $2\,\text{mW}$. Without illumination, the current $I_{Rx,0}$ is negligibly small. For optical input powers of $P_{Rx,0}=1\,\text{mW}$ and $2\,\text{mW}$, the currents $I_{Rx,0}$ are zero at a voltage $U_{Rx,0}$ of 0.19 V and 0.21 V, respectively.

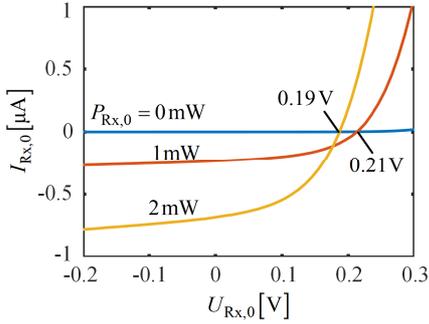

**Figure S3:** Estimation of build-up potential of an externally illuminated PIPED with zero bias-current. The static current-voltage characteristic is recorded without illumination, $P_{Rx,0}=0\,\text{mW}$, as well as for optical input powers of $1\,\text{mW}$ and $2\,\text{mW}$, measured in the silicon waveguide which feeds the PIPED. With an illumination of $P_{Rx,0}=1\,\text{mW}$ and $P_{Rx,0}=2\,\text{mW}$, the current $I_{Rx,0}$ vanishes at a voltage $U_{Rx,0}$ of 0.19 V and 0.21 V, respectively. These voltages correspond to the open-circuit voltages that appear with zero bias-current.

## 4. Equivalent-circuit of PIPED in T-wave systems

The schematic of an antenna-coupled Tx PIPED is shown in Fig. S4(a), where THz transmitter and receiver carrier are derived from the same optical sources (homodyne reception). The PIPED is directly connected to the antenna feed points and is forward biased by a DC voltage $U_{Tx,0}$, which is applied via a bias line, see Inset 1 in Fig. 1 in the main text. This bias line has a low-pass characteristic, which prevents leakage of the THz current to the DC voltage source $U_{Tx,0}$, and is represented by an inductor $L_0$. The PIPED is illuminated by the optical power $P_{Tx}(t)$ defined in Eq. (S2),

$$P_{Tx}(t) = P_{Tx,0}(t) + \hat{P}_{Tx,1}(t)\cos(\omega_{Tx,THz}t + \varphi_{Tx,THz}). \quad (S15)$$

We assume that only the amplitude $\hat{P}_{Tx,1}(t)$ of the optical power oscillation is modulated, but not its phase $\varphi_{Tx,THz}$. For analyzing the behavior of the system at THz frequencies, we translate the schematic into an equivalent-circuit representation using small-signal approximations. To this end, the PIPED current is represented by a low-frequency component $I_{Tx,0}(t)$, and a component with amplitude $\hat{I}_{Tx,1}(t)$ oscillating at THz frequency $\omega_{Tx,THz}$, see Eq. (S3),

$$\begin{aligned}I_{Tx}(t) &= I_{Tx,0}(t) + I_{Tx,1}(t) \\ &= I_{Tx,0}(t) + \hat{I}_{Tx,1}(t)\cos(\omega_{Tx,THz}t + \varphi_{Tx,THz}).\end{aligned} \quad (S16)$$

The PIPED voltage is represented in an analogous way,

$$\begin{aligned}U_{Tx}(t) &= U_{Tx,0}(t) + U_{Tx,1}(t) \\ &= U_{Tx,0}(t) + \hat{U}_{Tx,1}(t)\cos(\omega_{Tx,THz}t + \varphi_{Tx,THz}).\end{aligned} \quad (S17)$$

Using the small-signal approximation, the behavior of the Tx PIPED can then be modelled by a linearized voltage-dependent sensitivity,

$$S_{Tx}(U_{Tx}(t)) \approx S_{Tx}(U_{Tx,0}) + \left.\frac{dS_{Tx}}{dU_{Tx}}\right|_{U_{Tx,0}} U_{Tx,1}(t). \quad (S18)$$

The Tx PIPED current is then given by

$$\begin{aligned}I_{Tx}(t) &= S_{Tx}(U_{Tx}(t))P_{Tx}(t) \\ &= \underbrace{S_{Tx}(U_{Tx,0})P_{Tx,0}(t)}_{(1)} \\ &+ \underbrace{\tfrac{1}{2}\left.\frac{dS_{Tx}}{dU_{Tx}}\right|_{U_{Tx,0}}\hat{P}_{Tx,1}(t)\hat{U}_{Tx,1}(t)}_{(2)} \\ &+ \underbrace{S(U_{Tx,0})\hat{P}_{Tx,1}(t)\cos(\omega_{Tx,THz}t + \varphi_{Tx,THz})}_{(3)} \\ &+ \underbrace{\left.\frac{dS_{Tx}}{dU_{Tx}}\right|_{U_{Tx,0}} P_{Tx,0}(t)\hat{U}_{Tx,1}(t)\cos(\omega_{Tx,THz}t + \varphi_{Tx,THz})}_{(4)} \\ &+ \underbrace{\tfrac{1}{2}\left.\frac{dS_{Tx}}{dU_{Tx}}\right|_{U_{Tx,0}}\hat{P}_{Tx,1}(t)\hat{U}_{Tx,1}(t)\cos(2\omega_{Tx,THz}t + 2\varphi_{Tx,THz}).}_{(5)}\end{aligned}$$
(S19)

Note that the time dependence of voltages, currents and powers $U_{Tx,0}(t)$, $I_{Tx,0}(t)$, $P_{Tx,0}(t)$, $\hat{U}_{Tx,1}(t)$, $\hat{I}_{Tx,1}(t)$, $\hat{P}_{Tx,1}(t)$ result from an auxiliary low-frequency modulation, which could be imposed for lock-in detection. The quantities $P_{Tx,0}$ and $\hat{P}_{Tx,1}$ have the same order of magnitude. The first two expressions (1) and (2) in Eq. (S19) describe a low-frequency current. Because $S_{Tx} \gg (dS_{Tx}/dU_{Tx})\hat{U}_{Tx,1}$ holds, subexpression (2) can be neglected compared to subexpression (1), and subexpressions (4), (5) can be neglected compared to subexpression (3). This was the implicit approximation assumed in Eq. (S4).

From Eq. (S19), we derive the small-signal equivalent circuit at THz frequencies, see Fig. S4(b). The dominating THz current contribution (3) is modelled by a THz current source with a slowly varying complex amplitude,

$$\underline{I}_{Tx,1}(t) = S(U_{Tx,0})\hat{P}_{Tx,1}(t)e^{j\varphi_{Tx,THz}}. \quad (S20)$$

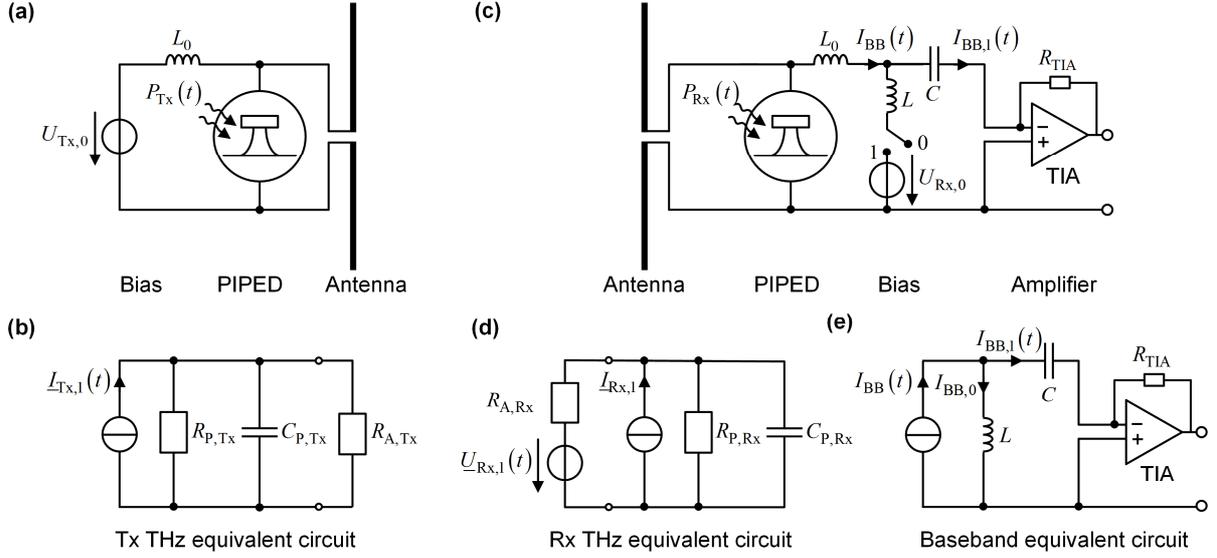

**Figure S4:** Equivalent circuit of PIPED in T-wave systems. (a) Schematic of antenna-coupled T-wave transmitter. The PIPED is directly connected to the antenna feed points and is biased by the voltage $U_{\text{Tx},0}$, which is applied via a bias transmission line acting as a low pass and represented by an inductor $L_0$. The PIPED is driven by the optical power $P_{\text{Tx}}(t)$. (b) Small-signal equivalent-circuit diagram of the Tx at THz frequencies. The PIPED is represented by the THz current source with complex amplitude $\underline{I}_{\text{Tx},1}(t) = S(U_{\text{Tx},0})\hat{P}_{\text{Tx},1}(t)\exp[j\varphi_{\text{Tx,THz}}(t)]$, the internal resistor $R_{\text{P,Tx}}$ as well as by the capacitor $C_{\text{P,Tx}}$. The antenna is modelled by the radiation resistor $R_{\text{P,Tx}}$. (c) Schematic of an antenna coupled to a T-wave receiver. The PIPED is directly connected to the antenna feed points and is biased via transmission lines acting as a low pass and represented by an inductor $L_0$. The receiver is either operated with an externally applied voltage $U_{\text{Rx},0}$ (switch position 1), or with a zero bias current (switch position 0). A bias-T, represented by the capacitor $C$ and the inductor $L$, extracts the time-varying part $I_{\text{BB},1}(t)$ of the total baseband current $I_{\text{BB}}(t)$. A transimpedance amplifier (TIA) with transimpedance $R_{\text{TIA}}$ amplifies the baseband current $I_{\text{BB},1}(t)$ and feeds it to a lock-in amplifier (not depicted). (d) Small-signal equivalent circuit of the Rx at THz frequencies. The PIPED is modelled by the THz current source with complex amplitude $\underline{I}_{\text{Rx},1} = S(U_{\text{Rx},0})\hat{P}_{\text{Rx},1}\exp[j\varphi_{\text{Rx,THz}}(t)]$, an internal resistor $R_{\text{P,Rx}}$ and a capacitor $C_{\text{P,Rx}}$, similarly as with the Tx shown in Subfigure (b). The receiving antenna is modelled by a voltage source $\underline{U}_{\text{Rx},1}(t) = \hat{U}_{\text{Rx},1}(t)\exp[j(\varphi_{\text{Tx,THz}} + \varphi_{\text{TxRx}})]$ and by a radiation resistor $R_{\text{P,Rx}}$. In our experiments, the THz signal and therefore $\hat{U}_{\text{Rx},1}$ are low-frequency modulated for lock-in detection. (e) Baseband equivalent circuit. The down-conversion leads to a baseband current $I_{\text{BB}}(t)$, see Eq. (S24). The bias-T branches this current in a constant part $I_{\text{BB},0}$ and in a time-varying part $I_{\text{BB},1}(t)$. The time-varying current component is fed to the TIA.

Note that the THz current source is ideal meaning that it features an internal conductance of zero. For a more realistic representation, it may be useful to also consider the impact of a finite source conductance. This is equivalent to considering expression (4) in Eq. (S19), which describes a contribution to the THz current amplitude $\hat{I}_{\text{Tx},1}(t)$ that increases in proportion to the THz voltage amplitude $\hat{U}_{\text{Tx},1}(t)$. In the equivalent-circuit diagram, this contribution is represented by the PIPED resistance

$$R_{\text{P,Tx}} = \dfrac{1}{\left.\dfrac{dS_{\text{Tx}}}{dU_{\text{Tx}}}\right|_{U_{\text{Tx},0}} P_{\text{Tx},0}} = \left.\dfrac{dU_{\text{Tx},0}}{dI_{\text{Tx},0}}\right|_{U_{\text{Tx},0},P_{\text{Tx},0}}. \quad (S21)$$

Furthermore, we introduce the capacitance $C_{\text{P,Tx}}$ into the THz equivalent circuit diagram to account for the tiny, but nonzero capacitance of the metal-coated sidewalls of the PIPED, see Fig. 2(a) in the main text. The antenna is modelled by its radiation resistance $R_{\text{A,Tx}}$.

In an analogous way we describe the PIPED Rx. Figure S4(c) shows the schematic of the Rx PIPED. Again, the PIPED is directly connected to the antenna feed points, and the bias line is represented by the inductor $L_0$. As described in Section 3, the Rx PIPED can either be operated with zero bias current if the switch is set to position 0, or with an externally applied bias voltage $U_{\text{Rx},0}$ if the switch is set to position 1. A bias-T, represented by the inductor $L$ and the capacitor $C$, separates the time varying part $I_{\text{BB},1}(t)$ of the baseband current $I_{\text{BB}}(t)$ from the constant part. The current $I_{\text{BB},1}(t)$ is amplified in a transimpedance amplifier with a transimpedance of $R_{\text{TIA}} = 1\text{M}\Omega$.

To obtain the THz small-signal equivalent circuit of the Rx, we follow the same procedure as with the Tx PIPED. This leads to the equivalent circuit shown in Fig S4(d). In this case, the Rx antenna and the received THz wave are modelled by a radiation resistor $R_{\text{A,Rx}}$ and a voltage source with a slowly varying complex amplitude $\underline{U}_{\text{Rx},1}(t)$, see Eq. (S8),

$$\underline{U}_{\text{Rx},1}(t) = \hat{U}_{\text{Rx},1}(t) e^{j(\varphi_{\text{Tx,THz}} + \varphi_{\text{TxRx}})}. \quad (S22)$$

The time-dependence of $\hat{U}_{\text{Rx},1}(t)$ originates from a modulation of the Tx DC bias voltage $U_{\text{Tx},0}$, which leads to a varying THz signal amplitude. In contrast, the THz current source

$$\underline{I}_{\text{Rx},1} = S(U_{\text{Rx},0})\hat{P}_{\text{Rx},1} e^{j\varphi_{\text{Rx,THz}}} \quad (S23)$$

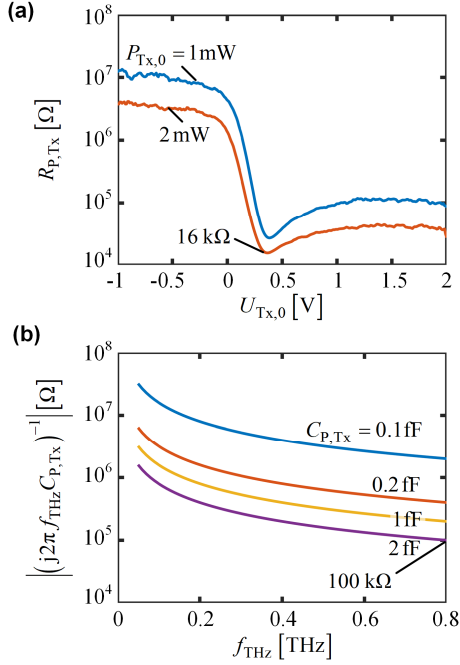

**Figure S5:** Estimation of Tx PIPED impedance **(a)** PIPED resistance $R_{P,Tx}$ estimated by taking the different resistance $dU_{Tx,0}/dI_{Tx,0}|_{P_{Tx,0}}$ of the static voltage-current characteristic for different optical input powers $P_{Tx,0}$. For all operation conditions, the differential resistance is larger than 10 kΩ. **(b)** Magnitude of the impedance $(j2\pi f_{THz} C_{P,Tx})^{-1}$ associated with the PIPED capacitance $C_{P,Tx}$. In the frequency of interest, the impedance magnitude is larger than 100 kΩ.

is unmodulated.

The phase and the amplitude of the received THz wave can be extracted from the time-dependent baseband current $I_{BB}(t)$, see Eqs. (S12) and (S13),

$$I_{BB}(t) = I_{BB,0} + I_{BB,1}(t) \\ = I_{BB,0} + \hat{I}_{BB,1}(t)\cos(\varphi_{BB}),$$ (S24)

with the abbreviations

$$I_{BB,0} = S_{Rx}(U_{Rx,0})P_{Rx,0},$$

$$\hat{I}_{BB,1}(t) = \frac{1}{2}\frac{dS_{Rx}}{dU_{Rx}}\bigg|_{U_{Rx,0}} \hat{P}_{Rx,1}\hat{U}_{Rx,1}(t),$$ (S25)

$$\varphi_{BB} = \varphi_{Tx,THz} - \varphi_{Rx,THz} + \varphi_{TxRx}.$$

Note that the Rx PIPED current can be described in analogy to Eq. (S19) and that the quantities $I_{BB,0}$ and $I_{BB,1}(t)$ would correspond to the subexpressions (1) and (2). The time dependence of $I_{BB}(t)$ originates from the modulation of the Tx bias voltage $U_{Tx,0}$, which leads to a modulation of the received THz voltage amplitude $\hat{U}_{Rx,1}(t)$. This modulation allows to separate $\hat{I}_{BB,1}(t)$ from the constant current $I_{BB,0}$. The equivalent baseband circuit of the Rx is shown in Fig. S4(e). The time-varying current $I_{BB,1}(t)$ is separated from $I_{BB}(t)$ by the bias-T, which is modelled by the capacitor $C$ and the inductor $L$. To this end, the high-pass cut-off frequency of the bias-T must be smaller than the modulation frequency for lock-in reception. The baseband current $I_{BB,1}(t)$, which is in proportion to the incoming THz field, is amplified by the transimpedance amplifier, and fed to a subsequent lock-in amplifier.

Since the Tx PIPED features a large internal resistance $R_{P,Tx}$ and a small capacitance $C_{P,Tx}$, the corresponding components in the equivalent-circuit representation can often be neglected. To quantify the PIPED resistance, we calculate the differential resistance $dU_{Tx,0}/dI_{Tx,0}|_{P_{Tx,0}}$ from the static voltage-current characteristics measured for different optical input powers $P_{Tx,0}$, see Fig. S5(a) and Eq. (S21). For all operating conditions, we find that the differential resistance is much larger than 10 kΩ. In addition, the small footprint of the PIPED leads to capacitances[1] well below 2 fF. Figure S5(b) shows the magnitude of the frequency-dependent impedance associated with this capacitance. In the frequency range of interest, the magnitude of the impedance is larger than 100 kΩ. Equivalent statements hold for the Rx PIPED. These large internal impedances of the PIPED allow to neglect $R_{P,Tx,Rx}$ in comparison to the antenna radiation resistance $R_{A,Tx,Rx}$ which is typically in the order of 100 Ω.

Figure S6(a) illustrates the equivalent circuit of the monolithically integrated T-wave system, see Figure 4 in the main text. The Tx PIPED and the Rx PIPED are modelled as described in Fig. S4(b) and (d). The transmission line is represented by a complex impedance matrix with components $\underline{Z}_{11}$, $\underline{Z}_{21} = \underline{Z}_{12}$ and $\underline{Z}_{22}$ of a reciprocal two-port network[2]. Making use of the fact that the PIPED impedance is much larger than the impedance matrix elements $\underline{Z}_{ij}$, we simplify the equivalent circuit by eliminating the internal resistors and capacitors of the Rx and Tx PIPED.

Further simplifications are possible, if we modulate the bias $U_{Tx,0}$ of the Tx PIPED and hence the Tx current amplitude $\underline{I}_{Tx,1}(t)$, and exploit lock-in detection of the baseband current $I_{BB}(t)$. Therefore, we only extract the component $I_{BB,1}(t)$ that is related to the THz voltage $U_{Rx,1}(t)$ received from the transmitter. The THz current $\underline{I}_{Rx,1}$ is not affected by this modulation, and hence does not influence the lock-in signal. The current source $\underline{I}_{Rx,1}$ can therefore be omitted for the further analysis of the equivalent circuit. This leads to the simplified diagram shown in Fig S6(b) and to the relation

$$\underline{U}_{Rx,1}(t) = \underline{Z}_{21}\underline{I}_{Tx,1}(t).$$ (S26)

Using this relation we can derive a connection of the complex transfer impedance $\underline{Z}_{21}$ with the baseband current $I_{BB,1}(t)$. With Eq. (S20), Eq. (S22), and the definition of the transfer impedance $\underline{Z}_{21} = |\underline{Z}_{21}|e^{j\varphi_{21}}$, Eq. (S26) is written as

$$\hat{U}_{Rx,1}(t)e^{j(\varphi_{Tx,THz}+\varphi_{TxRx})} = |\underline{Z}_{21}|e^{j\varphi_{21}}\hat{I}_{Tx,1}e^{j\varphi_{Tx,THz}}.$$ (S27)

The relation can be split into one equation for the magnitude $|\underline{Z}_{21}|$ and another equation for the phase $\varphi_{21}$ of the transfer impedance,

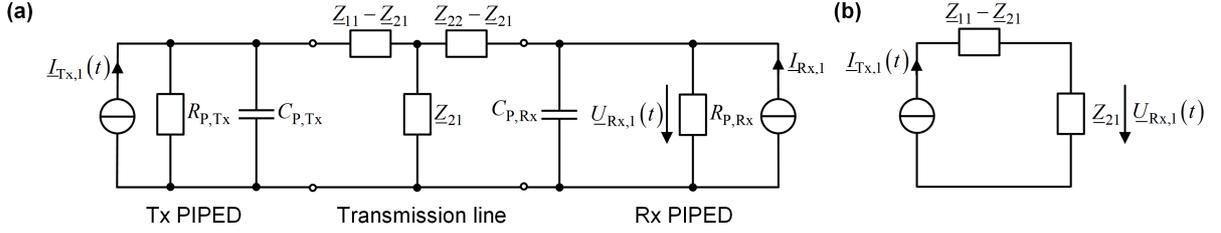

**Figure S6:** Equivalent circuit of the monolithically integrated T-wave system. **(a)** Small-signal equivalent circuit. The PIPED Tx and PIPED Rx are modelled as shown in Fig. S4(b) and (d). The transmission line is described by an impedance matrix with elements $\underline{Z}_{11}$, $\underline{Z}_{22}$ and $\underline{Z}_{21} = \underline{Z}_{12}$ for a reciprocal two-port network. **(b)** Simplified equivalent circuit, considering only quantities that are modulated by the lock-in signal of the Tx PIPED bias $U_{\mathrm{Tx},0}(t)$, and exploiting the fact that the PIPED impedance is much larger than the transmission line impedances.

$$\hat{U}_{\mathrm{Rx},1}(t) = |\underline{Z}_{21}|\hat{I}_{\mathrm{Tx},1}(t), \qquad \text{(S28)}$$
$$\varphi_{21} = \varphi_{\mathrm{TxRx}}.$$

Substituting Eq. (S28) into Eq. (S25) leads to

$$\hat{I}_{\mathrm{BB},1}(t) = \tfrac{1}{2} \left.\frac{\mathrm{d}S_{\mathrm{Rx}}}{\mathrm{d}U_{\mathrm{Rx}}}\right|_{U_{\mathrm{Rx},0}} \hat{P}_{\mathrm{Rx},1} \hat{I}_{\mathrm{Tx},1}(t)|\underline{Z}_{21}|, \qquad \text{(S29)}$$
$$\varphi_{\mathrm{BB}} = \varphi_{\mathrm{Tx,THz}} - \varphi_{\mathrm{Rx,THz}} + \varphi_{21}.$$

In this equation the Tx current $\hat{I}_{\mathrm{Tx},1}(t) = \hat{I}_{\mathrm{Tx},1}\cos(\omega_{\mathrm{LI}} t)$ with amplitude $\hat{I}_{\mathrm{Tx},1}$, is subject to a sinusoidal modulation with the lock-in frequency $\omega_{\mathrm{LI}}$. The baseband photocurrent $\hat{I}_{\mathrm{BB},1}(t) = \hat{I}_{\mathrm{BB},1}\cos(\omega_{\mathrm{LI}} t)$ with amplitude $\hat{I}_{\mathrm{BB},1}$ can then be used as a measure for the magnitude of the transfer impedance $|\underline{Z}_{21}|$. Substituting $\hat{I}_{\mathrm{Tx},1}(t) = S_{\mathrm{Tx}}(U_{\mathrm{Tx},0})\hat{P}_{\mathrm{Tx},1}$ we write

$$\begin{aligned}\hat{I}_{\mathrm{BB},1} &= \tfrac{1}{2} \left.\frac{\mathrm{d}S_{\mathrm{Rx}}}{\mathrm{d}U_{\mathrm{Rx}}}\right|_{U_{\mathrm{Rx},0}} \hat{P}_{\mathrm{Rx},1} \hat{I}_{\mathrm{Tx},1}|\underline{Z}_{21}| \\ &= \tfrac{1}{2} \left.\frac{\mathrm{d}S_{\mathrm{Rx}}}{\mathrm{d}U_{\mathrm{Rx}}}\right|_{U_{\mathrm{Rx},0}} \hat{P}_{\mathrm{Rx},1} S_{\mathrm{Tx}}(U_{\mathrm{Tx},0}) \hat{P}_{\mathrm{Tx},1}|\underline{Z}_{21}| \quad \text{(S30)} \\ &= |\underline{Z}_{21}|/a.\end{aligned}$$

In this relation, the proportionality factor $1/a$ is given by

$$\frac{1}{a} = \tfrac{1}{2} \left.\frac{\mathrm{d}S_{\mathrm{Rx}}}{\mathrm{d}U_{\mathrm{Rx}}}\right|_{U_{\mathrm{Rx},0}} \hat{P}_{\mathrm{Rx},1} S_{\mathrm{Tx}}(U_{\mathrm{Tx},0}) \hat{P}_{\mathrm{Tx},1}, \qquad \text{(S31)}$$

and has the unit $[\mathrm{A}/\Omega]$. As a result, we can determine $|\underline{Z}_{21}|$ by measuring $\hat{I}_{\mathrm{BB},1}$. In addition, the phase $\varphi_{21}$ of the transfer impedance $\underline{Z}_{21}$ can be determined by measuring the phase $\varphi_{\mathrm{BB}}$ of the baseband current in Eq. (S29). We eliminate the influence of the unknown group delay in the optical fibers, see the Methods section of the main text, by swapping the role of the Tx and the Rx PIPED and recording $\varphi_{\mathrm{BB,rev}} = -(\varphi_{\mathrm{Tx,THz}} - \varphi_{\mathrm{Rx,THz}}) + \varphi_{21}$. By adding Eq. (S29) we finally find $\varphi_{21}$.

### 5. Frequency roll-off of PIPED

For large frequencies, the conversion factor of the Tx PIPED and of the Rx PIPED drop due to limitations by the carrier transit time, see Section "Plasmonic internal photoemission detectors (PIPED) for optoelectronic T-wave processing" of the main text. In the following, we characterize the high-speed performance and the frequency roll-off of the PIPED. The frequency response of the Tx and Rx PIPED photocurrent

$$I_{\mathrm{Tx,Rx}}(f_{\mathrm{THz}}) = \eta_{\mathrm{Tx,Rx}}(f_{\mathrm{THz}}) S_{\mathrm{Tx,Rx}} P_{\mathrm{Tx,Rx}} \qquad \text{(S32)}$$

is characterized an RC low pass characteristic,

$$\eta_{\mathrm{Tx}}(f_{\mathrm{THz}}) = \eta_{\mathrm{Rx}}(f_{\mathrm{THz}}) = \frac{1}{\sqrt{1 + (f_{\mathrm{THz}}/f_{3\mathrm{dB}})^2}}. \qquad \text{(S33)}$$

Figure S7(a) shows the simulated magnitude $|\underline{Z}_{21,\mathrm{sim}}(f_{\mathrm{THz}})|$ of the transfer impedance for the $L = 0.01\,\mathrm{mm}$ long transmission line together with the measured frequency-dependent amplitude $\hat{I}_{\mathrm{BB},1}(f_{\mathrm{THz}})$ of the baseband photocurrent. For an ideal PIPED, $\hat{I}_{\mathrm{BB},1}(f_{\mathrm{THz}}) = |\underline{Z}_{21}(f_{\mathrm{THz}})|/a$ would be true, see Eq. (S31). The real device shows a low pass characteristic. To match the simulation to the measurements, we include the RC low pass characteristic of Eq. (S33) and find

$$\begin{aligned}\hat{I}_{\mathrm{BB},1}(f_{\mathrm{THz}}) &= \eta_{\mathrm{Tx}}(f_{\mathrm{THz}})\eta_{\mathrm{Rx}}(f_{\mathrm{THz}})|\underline{Z}_{21,\mathrm{sim}}(f_{\mathrm{THz}})|/a \\ &= \frac{1}{1 + (f_{\mathrm{THz}}/f_{3\mathrm{dB}})^2}|\underline{Z}_{21,\mathrm{sim}}(f_{\mathrm{THz}})|/a.\end{aligned} \qquad \text{(S34)}$$

The parameters $a$ and $f_{3\mathrm{dB}}$ are obtained from the measurement data by a least-squares fit. This is necessary because the experimental operating conditions of the Tx and Rx PIPED at THz frequencies are only approximately known. We obtain $a = 21\,\Omega/\mathrm{nA}$ and $f_{3\mathrm{dB}} = 0.3\,\mathrm{THz}$, indicating again the superior bandwidth properties of the PIPED. Figure S7(b) shows the double-logarithmic display of the PIPED frequency roll-off. We attribute the same 3 dB bandwidth to various PIPED for all measuements in Fig. 4(c)-(e) in the main text. Residual differences between the PIPED employed for measuring with different lengths of the transmission line ($L = 0.63\,\mathrm{mm}$ and $L = 0.95\,\mathrm{mm}$) we compensate by fitting $a$.

Using the parameters $a$ and $f_{3\mathrm{dB}}$, we translate the measured photocurrent amplitude $\hat{I}_{\mathrm{BB},1}(f_{\mathrm{THz}})$ into the corresponding transfer impedance $|\underline{Z}_{21,\mathrm{meas}}(f_{\mathrm{THz}})|$, which is plotted along with its simulated counterpart in Fig. S7(c); this corresponds to Fig. 4(c) of the main text. Simulation and measurement show good agreement.

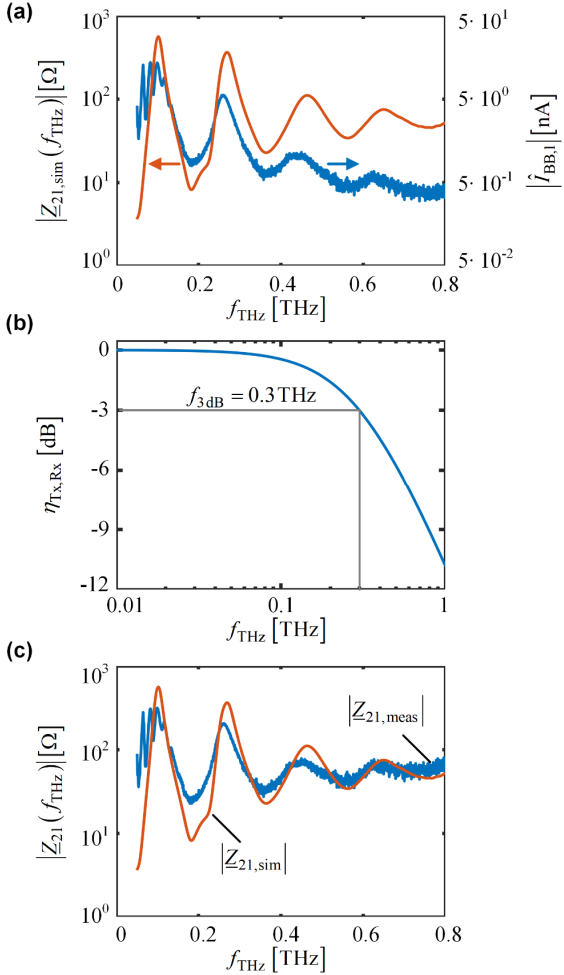

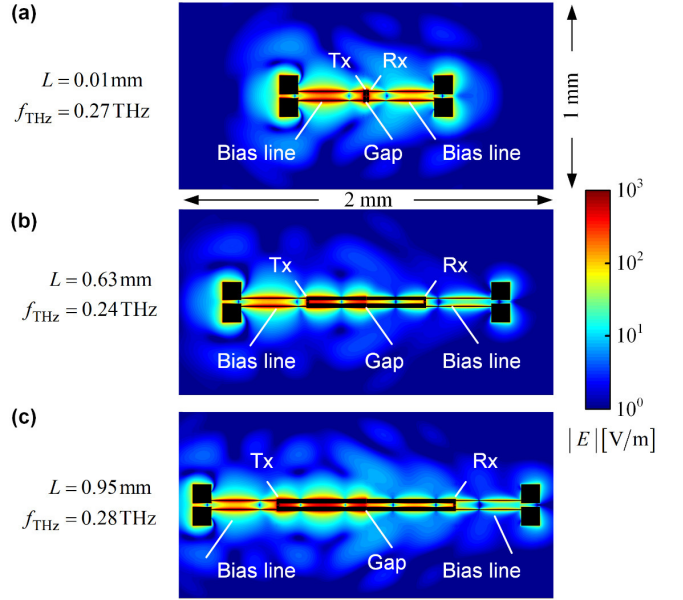

**Figure S7:** Characterization of PIPED frequency roll-off frequency $f_{3dB}$. **(a)** Comparison of simulated magnitude $|Z_{21,\text{sim}}(f_{\text{THz}})|$ of the transfer impedance (red, left axis) and the measured amplitude $\hat{I}_{\text{BB,1}}(f_{\text{THz}})$ of the baseband photocurrent (blue, right axis) obtained for a transmission line length $L = 0.01\,\text{mm}$. The limited PIPED bandwidth causes a decrease of the current amplitude $\hat{I}_{\text{BB,1}}(f_{\text{THz}})$ with respect to the simulated magnitude $|Z_{21,\text{sim}}(f_{\text{THz}})|$ at larger frequencies. **(b)** PIPED Tx and PIPED Rx frequency response approximated by an RC low pass. The parameter $f_{3dB} = 0.3\,\text{THz}$ was obtained by matching in subfigure (a) the two traces. **(c)** Simulated and measured transfer impedance for a transmission line length $L = 0.01\,\text{mm}$. The measured photocurrent amplitudes $\hat{I}_{\text{BB,1}}(f_{\text{THz}})$ are expressed by the modulus corresponding transfer impedances $|Z_{21,\text{meas}}(f_{\text{THz}})|$ using the fitting parameters $a = 21\,\Omega/\text{nA}$ and $f_{3dB} = 0.3\,\text{THz}$.

## 6. Electric field distribution of integrated T-wave system

The transfer impedances of the various THz transmission lines plotted in Fig. 4(c)-(e) of the main text exhibit a resonant behavior. For a better understanding of these resonances, we perform for selected frequencies simulations of the electric field distribution along the transmission line. Figures S8(a)-(c) display the simulation results for the magnitude of the electric field obtained for three different transmission line lengths $L = (0.01, 0.63, 0.95)\,\text{mm}$. The PIPED Tx is modelled as a THz current source with a magnitude chosen to be $50\,\mu\text{A}$, corresponding to the DC current generated in the PIPED, see Methods. The frequencies $f_{\text{THz}} = (0.27, 0.24, 0.28)\,\text{THz}$ represent maxima of the transfer impedance magnitude $|Z_{21,\text{sim}}|$ in Fig. 4(c)-(e). The field simulations exhibit standing wave patterns for the various lengths. We further find that the PIPED bias lines, which have a geometrical length of $0.35\,\text{mm}$, influence the resonance frequencies of the transmission line. Thinner and longer bias lines could reduce this influence.

**Figure S8:** Simulated magnitude of the electric field for the resonance frequencies $f_{\text{THz}} = (0.27, 0.24, 0.28)\,\text{THz}$ at the (geometrical) transmission line lengths $L = (0.01, 0.63, 0.95)\,\text{mm}$ in subfigures **(a)-(c)**. The bias lines have a length of $0.35\,\text{mm}$ each. The Tx is modelled as a THz current source with a current magnitude of $50\,\mu\text{A}$.

## 7. Conversion from S-parameters to Z-parameters

The complex transfer impedance $\underline{Z}_{21}$ of the THz resonator can be derived from numerically calculated S-parameters, see Methods of the main text. To this end, we use the following relations[2]:

$$\begin{aligned}
\underline{Z}_{11} &= Z_0 \frac{(1+\underline{S}_{11})(1-\underline{S}_{22}) + \underline{S}_{12}\underline{S}_{21}}{(1-\underline{S}_{11})(1-\underline{S}_{22}) - \underline{S}_{12}\underline{S}_{21}} \\
\underline{Z}_{12} &= Z_0 \frac{2\underline{S}_{12}}{(1-\underline{S}_{11})(1-\underline{S}_{22}) - \underline{S}_{12}\underline{S}_{21}} \\
\underline{Z}_{21} &= Z_0 \frac{2\underline{S}_{21}}{(1-\underline{S}_{11})(1-\underline{S}_{22}) - \underline{S}_{12}\underline{S}_{21}} \\
\underline{Z}_{22} &= Z_0 \frac{(1-\underline{S}_{11})(1+\underline{S}_{22}) + \underline{S}_{12}\underline{S}_{21}}{(1-\underline{S}_{11})(1-\underline{S}_{22}) - \underline{S}_{12}\underline{S}_{21}}
\end{aligned} \quad (S35)$$

## References

1. Muehlbrandt, S. *et al.* Silicon-plasmonic internal-photoemission detector for 40 Gbit/s data reception. *Optica* **3,** 741 (2016).
2. Pozar, D. M. *Microwave engineering.* (John Wiley & Sons, 2012).